\documentclass{nature}

\usepackage{graphicx}
\usepackage{multirow}
\graphicspath{ {images/} }
\DeclareGraphicsExtensions{.png,.pdf,.eps}

\usepackage{hhline}
\usepackage{epstopdf}
\usepackage{amsmath,amssymb}

\usepackage{comment}

\usepackage{xcolor}

\title{The coevolution of migrating planets and their pulsating stars through episodic resonance locking.}

\author{Jared Bryan$^{1}$, Julien de Wit$^{1}$, Meng Sun$^{2,3}$, Zo\"e L. de Beurs$^{1}$, Richard H. D. Townsend$^{2}$} 

\begin{document}
\maketitle
\begin{affiliations}
 \item Department of Earth, Atmospheric and Planetary Sciences, Massachusetts Institute of Technology, Cambridge, MA, USA;
 \item Department of Astronomy, University of Wisconsin -- Madison, Madison, WI, USA;
 \item Center for Interdisciplinary Exploration and Research in Astrophysics, Department of Physics \& Astronomy, Northwestern University, Evanston, IL, USA
\end{affiliations}

\begin{abstract}
Hot Jupiters are expected to form far from their host star and move toward close-in, circular orbits via a smooth, monotonic decay due to mild and constant tidal dissipation. 
Yet, systems exhibiting planet-induced stellar pulsations have recently been found, suggesting unexpectedly strong tidal interactions. 
Here we combine stellar evolution and tide models to show that dynamical tides raised by eccentric gas giants can excite chains of resonance locks with multiple modes, enriching the dynamics seen in single-mode resonance locking of circularized systems. 
These series of resonance locks yield orders-of-magnitude larger changes in eccentricity and harmonic pulsations relative to those expected from a single episode of resonance locking or nonresonant tidal interactions. 
Resonances become more frequent as a star evolves off the main sequence providing an alternative explanation to the origin of some stellar pulsators and yielding the concept of ``dormant migrating giants''. 
Evolution trajectories are characterized by competing episodes of inward/outward migration and spin-up/-down of the star which are sensitive to the system parameters, revealing a new challenge in modeling migration paths and in contextualizing the observed populations of giant exoplanets and stellar binaries. 
This sensitivity however offers a new window to constrain the stellar properties of planetary hosts via tidal asteroseismology.
\end{abstract}

\flushbottom
\maketitle

\thispagestyle{empty}
\section{Main}
Hot Jupiters were the first exoplanets found around main sequence stars\cite{mayor1995jupiter}, but have no analog in our solar system.
Understanding the formation channels of hot Jupiters provides a strong test for planetary formation theories. 
Although it has been suggested that Hot Jupiters can form in situ\cite{batygin2016situ}, their formation beyond the ice line is favored, followed by inward migration. 
Inward migration proceeds either by interaction with the protoplanetary disk\cite{goldreich1980disk,lin1986tidal} or by a combination of eccentricity excitation (e.g. by planet-planet scattering\cite{rasio1996dynamical} or by cyclic\cite{kozai1962secular, lidov1962evolution} or chaotic secular interactions\cite{wu2011secular}) followed by tidal dissipation\cite{barker2009tidal, dawson2018origins} in the planet\cite{eggleton1998equilibrium} or its host star\cite{jackson2008tidal}.
Abundant evidence for tidal interactions exist for heartbeat stellar binaries in the form of photometric observations of tidally excited oscillations\cite{fuller2017accelerated, hambleton2018kic}.
High-amplitude stellar pulsations and rapid orbital evolution have appeared as evidence for unexpectedly strong tidal interaction in the 2.6~Gyrs-old\cite{bonomo2017gaps} system hosting the eccentric (e$\sim$0.52) hot Jupiter HAT-P-2b\cite{bakos2007hd, de2017planet, de2023revisiting}, with similar high-amplitude stellar pulsations in the WASP-33\cite{herrero2011wasp, kalman2022gravity} and HD-31221\cite{kalman2023discovery} systems.
Unexpectedly efficient energy and angular momentum transfer between a hot Jupiter and its host may provide new insights into their observed population and also yield frequent planetary engulfment as recently observed in the ZTF SLRN-2020 system\cite{de2023engulf}. 
Prompted by these systems, we investigate the tidal evolution of eccentric hot Jupiters.

Tidal migration of hot Jupiters due to stellar tides occurs through damping of large-scale tidal distortion (the equilibrium tide) by turbulent viscosity in the convective envelope\cite{zahn1977tidal, hut1981tidal, eggleton1998equilibrium}, as well as by radiative damping of dynamically excited internal gravity waves (the dynamical tide)\cite{cowling1941non, zahn1977tidal, goodman1998dynamical}. 
Damping of inertial waves\cite{ogilvie2007tidal, barker2022tidal} and nonlinear damping due to wave breaking near the stellar center\cite{weinberg2012nonlinear, weinberg2017tidal} or stellar surface\cite{macleod2022tidal} have also been considered by previous work.

Most previous work parameterized the stellar response to a tidal perturber in terms of a constant tidal quality factor\cite{goldreich1966q}, or a frequency-averaged tidal quality factor\cite{ogilvie2004tidal}. 
Since more sophisticated models of tidal evolution\cite{witte1999tidal, sun2023tides} demonstrate that orbital evolution rates are a sensitive function of the forcing frequency and the stellar structure, these models are insufficient to capture the dynamics of tidal evolution.

Previous work has also considered tides to be a purely dissipative process. 
Some authors have studied inverse tides due to the interaction between tidally forced oscillations and self-excited stellar oscillations, though only for circular stellar binaries\cite{willems2003nonadiabatic, fuller2021inverse}.
Inverse tides arising from Doppler shifting of stellar modes due to stellar rotation have also been studied for solar-type stars\cite{witte2002orbital}.
Another thread of research has dealt with chaotic tidal evolution for high-eccentricity ($e\geq0.95$) exoplanets where energy can be traded between the planetary f-mode and the orbital angular momentum\cite{ivanov2004tidal, wu2018diffusive, veras2019tidal, vick2019chaotic}. 
The oscillation of specific orbital elements during single resonances have also been explored, including the orbital inclination\cite{millholland2019obliquity, alexander2023orbital} through obliquity tides. 
In general, these resonances are modeled over short timescales, similar to applications considering stellar mode amplitude through dynamical resonance locking\cite{burkart2014dynamical}.

Much work on high eccentricity tidal migration considers only the effect of tides within the planet, neglecting tides in the star\cite{goldreich1966q,ogilvie2004tidal,vick2019chaotic}.
We consider dynamically excited g-modes in stars.
These can be resonantly excited, resulting in orders of magnitude higher pulsation amplitudes and rates of energy and angular momentum exchange between the orbit and the star than is possible with equilibrium tides alone\cite{witte2002orbital,ma2021orbital}.

It is also rare to see coupled stellar and orbital evolution, precluding the existence of resonance locking.
Resonance locking occurs when the tidal forcing frequency and the frequency of a stellar pulsation mode vary in concert, allowing for resonant interactions to be sustained over longer timescales than if either of these frequencies were held constant.
Resonance locking has been studied in binary stars\cite{witte1999tidal, fuller2012dynamical, burkart2012tidal, burkart2013tidal, burkart2014dynamical}.
Observational evidence comes from high-amplitude tidally excited oscillations in heartbeat stars\cite{fuller2017accelerated, hambleton2018kic}.
Recent work examined resonance locking as a mechanism for rapid orbital migration of a massive planet on a circular orbit\cite{ma2021orbital}.
It has also been proposed to drive orbital migration of massive exoplanets on eccentric orbits around solar-type stars\cite{witte2002orbital}, though nonlinear tidal dissipation due to wave breaking in the radiative core may prevent the application of resonance locking to the most massive exoplanets\cite{barker2010internal, ma2021orbital}.
It may also drive the orbital evolution of Saturn's and Jupiter's moons\cite{fuller2016resonance, lainey2020resonance}.
However, resonance locking of massive exoplanets on eccentric orbits around main sequence stars has received less attention, and it remains unclear how the enriched tidal forcing spectrum affects their tidal evolution.

We build on these studies by examining coupled stellar and orbital evolution for gas giants on eccentric, spin-aligned orbits around F-type main sequence stars.
We use the MESA stellar evolution code\cite{paxton2010modules,paxton2013modules,paxton2015modules,paxton2018modules,paxton2019modules,jermyn2023modules} to simulate the structural evolution of main sequence stars.
We focus on main sequence stars for which radiative damping of the dynamical tide is the dominant damping mechanism due to the large radiative envelope (Fig. 1).
These stars lack a deep convective envelope, so we neglect magnetic braking of the stellar rotation\cite{gossage2023magnetic}.
We then use the GYRE stellar pulsation code\cite{townsend2013gyre,townsend2018angular,goldstein2020contour,sun2023tides} to directly solve for the response of these stellar models to periodic tidal forcing by an eccentric point-mass companion.
In particular, we use GYRE-tides\cite{sun2023tides} to solve for the instantaneous rate-change of the orbital eccentricity, orbital semi-major axis, and stellar spin rate.
Beginning with an initial stellar and orbital configuration similar to that of HAT-P-2b\cite{bonomo2017gaps,stassun2019revised}, we integrate the orbital configuration and stellar spin state forward and backward in time.
In contrast to models of hot Jupiters on circular orbits, we demonstrate how orbital eccentricity greatly enriches the tidal forcing spectrum (Fig. 1).
In particular, we examine how competing resonances between different modes can lead to a wide array of tidal evolution behaviors, including brief resonance sweeps, sustained resonance locking, and wandering tidal migration.


\section{Results}
\subsection{Episodic and wandering tidal migration.}
Periods of slow orbital evolution are punctuated by rapid changes as the tidal forcing sweeps through resonance with a stellar eigenmode.
The equilibrium tidal response occurs at rates of $\dot{e}\sim10^{-11}$/yr, $\dot{a}\sim 10^{-12}$ au/yr and acts to circularize and shrink or expand the orbit while spinning up the star (Fig. 2a,b; Fig. 3).
This diversity in the senses of orbital evolution is possible due to the eccentric orbit interacting with multiple modes with different properties (Fig. 3).
However, resonances between the tidal forcing frequencies and natural stellar pulsation modes result in orders of magnitude more rapid orbital evolution, up to $\dot{e}\sim10^{-7}$/yr in the particular case shown in Fig. 2 and $\dot{e}\sim10^{-3}$/yr for an ensemble of stars shown in Fig. 4a.
Most of these resonances are exceedingly brief, lasting $\ll 1$ Myr at their peak.
However, when the rate of change of the frequency of the stellar pulsation mode due to stellar evolution and spin evolution matches the rate of change of the orbital forcing frequency, sustained excitation of individual stellar eigenmodes results in orbital evolution rates of $\dot{e}\sim10^{-8}-10^{-9}$/yrs over multi-Myr timescales, thereby driving the bulk of orbital evolution (Fig. 2, Fig. 4b).
This is a process known as resonance locking.

By contrast to planets on circular orbits these resonance locks do not persist over Gyr timescales\cite{ma2021orbital}.
This is because other modes compete with the resonantly locked mode, eventually pushing it out of resonance and resulting in relatively brief resonance locks ($\sim 1-5$ Myr).
However, the eccentric orbit enables a rich spectrum of possible resonance locks, decreasing the time between resonances.
In addition, the forest of possible resonances becomes more dense as the star evolves through the main sequence and we observe chains of resonance locks over timescales of tens of Myr (Fig. 2b inset).
Resonance locks can be seen as quasi plateaus in the orbital evolution rate (Fig. 2a,b insets).
This episodic tidal migration implies that observations of orbital evolution rates are generally not reflective of the long-term rates of orbital migration.
Instead, observations of high orbital evolution rates may be brief episodes in a migration history otherwise dominated by low tidal interaction strength.
Likewise, orbits not currently observed to be migrating cannot be assumed to be stationary over long timescales.

\subsection{Sensitivity of tidal evolution to initial conditions.}
The tidal evolution trajectory depends on the stellar and orbital configuration.
The array of stellar modes available for tidal interaction depends sensitively on the stellar structure.
In Fig. 4a, we show the rate-change in eccentricity for a range of stellar models between the zero-age main sequence and the red-edge of the main sequence for a fixed orbital configuration.
The stellar models range from $M=1.2-1.5 M_{\odot}$ with a constant metallicity of $Z=0.02$. 
No coherent patterns exist across the HR diagram due to the superposition of many different stellar modes which can contribute to the orbital evolution rate (patterns exist when considering one mode in isolation--see Methods; Extended Data Fig. 1, 2).
Without tuning the stellar models to fall exactly on resonances, we find orbital evolution rates up to $\dot{e}\sim 10^{-3}/yr$, far beyond the equilibrium tidal response of $\dot{e}\sim 10^{-11}$/yr.
In simulations with coevolving orbits and stellar structure, we observe $\dot{e}\sim10^{-4}$/yr (see Methods; Extended Data Fig. 3, 4).

To further investigate the sensitivity of the tidal evolution of massive exoplanets, we perform a sensitivity analysis of the tidal evolution trajectories for uncertainties in the initial orbital and stellar configuration.
To this end, we consider the parameters of the HAT-P-2 system\cite{bonomo2017gaps,stassun2019revised} and sample the $+1\sigma$, $-2\sigma$, and $+3\sigma$ in $a$, $e$, and $\Omega_{\mathrm{rot}}$, and $\pm 1\sigma$ in the $M$, $Z$, and $t_0$.
We then evolve the stellar structure and orbital configuration forward and backward in time and evaluate the similarity of the resulting orbital evolution trajectories.
We find that uncertainties in orbital parameters ($e$ and $a$) are small enough that orbital trajectories retain similar structure over $\sim 50$ Myr (Fig. 4b).
Similarities in the shape of the orbital trajectories imply similarities in the set of resonantly excited stellar modes.
The exact timing of these resonances differs on the order of $\sim 10$ Myr, far below the uncertainty in the stellar age and certainly below the main sequence lifetime of the star.
These small variations in the stellar structure and orbital configuration after a resonant tidal interaction cascade into larger changes for the next resonance, and the tidal evolution trajectories soon diverge.
Uncertainties in $t_0$, $\Omega_\mathrm{rot}$, $M$, and $Z$ immediately lead to qualitatively different tidal evolution behavior, both in the rates of orbital evolution and the specific timing of resonant interactions.
This sensitivity serves as a reminder of exoplanetary science's favorite tenet: ``Know thy star, know thy planet'' and highlights the challenge to our ability to robustly reconstruct the initial orbital configuration of hot Jupiters by time-reversing their tidal evolution. 
Reducing uncertainties in stellar parameters is thus critical for robust modeling of exoplanet tidal migration.

\section{Discussion}

\subsection{Challenges in reversing and propagating tidal evolution.}
Efforts to reconstruct the initial (prior to tidal migration) orbital configuration of hot Jupiters rely on a model of tidal migration.
Simplified theories based on a constant or frequency-averaged tidal quality factor lead to a smooth, monotonic evolution toward circular, synchronous, and short period orbits\cite{jackson2008tidal}.
Models based on sophisticated characterizations of the tidal response have tended to focus on the case of circular orbits\cite{ma2021orbital}, which again gives the impression of smooth, monotonic evolution of the orbit.
We find that orbital evolution is controlled by excitation of stellar g-modes, which are transiently excited to high amplitude and lock into resonance due to the coevolution of stellar modes, stellar spin rate, and orbital forcing frequencies leading to a strong sensitivity of the overall evolution trajectory to the system parameters.
Small perturbations to the orbital and stellar configuration cause the orbital trajectories to deviate above $1\sigma$ in the orbital parameters within 1-15 Myr (see methods; Extended Data Fig. 5, 6, 7). 

This sensitivity to the system parameters highlights a challenge to our ability to time-reverse and propagate the tidal evolution calculations in order to deterministically reconstruct the initial orbital configuration of hot Jupiters, a step that is key to contextualize the observed populations of hot Jupiters vs warm Jupiters, circularized vs eccentric ones.
Further work to investigate the statistics of tidal evolution trajectories would be needed to adequately account for and propagate the degree of confidence in the primitive orbital configurations of particular exoplanets onto their evolution trajectory.

This sensitivity to initial stellar and orbital configuration is also relevant to binary population synthesis studies.
In binary population synthesis, one of the goals is to determine the final mass and orbital distribution of compact objects, in order to aid in gravitational wave data analysis. 
To achieve this, an initial mass, orbital period, and mass ratio distribution are provided to launch large grids of binary evolution models. 
Due to computational constraints, numerical calculations of the tidal dissipation rate at each stellar evolutionary step, which require a stellar profile, are often replaced by fitting formulae and semi-analytical equations.
In this work, we demonstrate that tidal evolution in binaries is sensitive to parameters related to stellar and binary evolution, which implies that constructing more accurate parameterized equations for fast calculations could be problematic, particularly for the higher mass-ratio case of stellar binaries.

\subsection{The source of sign changes in the orbital evolution rates.}
The sensitivity of the trajectories also finds its origin in the sign changes in the orbital evolution rates. This sign of the orbital evolution rates depends on three quantities: (i) the sign of the Doppler-shifted forcing frequency, (ii) whether the position of the companion on-average lags or leads the location of the tidal disturbance for a particular mode, and (iii) the stability of the excited mode.
First, sign changes in the Doppler-shifted forcing frequency $\sigma_{m,k}=k\Omega_\mathrm{orb}-m\Omega_\mathrm{rot}$ determine whether the excited mode is prograde or retrograde in the corotating frame of the star.
Fig. 2 shows that a star rotating at $1.5\times$ the pseudosynchronous rotation rate\cite{hut1981tidal,sun2023tides,townsend2023discrepant} but otherwise with the orbital configuration of the HAT-P-2 system\cite{bonomo2017gaps} has $\sigma_{2,k}<0$ for $k<9$, and thus $\dot{e}>0$ (Fig. 2d). 
Higher rotation rates lead to additional retrograde modes, while lower rotation rates lead to only prograde modes (see Methods; Extended Data Fig. 3, 4).
Also visible in Fig. 2d is that $k=1,2$ retain the circularization sign of $\dot{e}$ ($\dot{e}<0$) despite being retrograde in the corotating frame.
This is because the tidal distortion for $m=2$, $k=1,2$ modes leads the position of the companion in an orbit-averaged sense, while for other $k$ it lags.
When combined with the negative Doppler-shifted forcing frequency, this results in modes that act to circularize the orbit.
The final source of sign changes in orbital evolution rates is due to oscillation mode instability such that the heat-engine of the star supplies energy to the planet's orbit, a phenomenon called inverse tides\cite{willems2003nonadiabatic,fuller2021inverse}.
The main sequence stars we study do not host unstable modes, so this mechanism is not applicable to this work.
Additional discussion of these mechanisms can be found in Sun et al. (2023).

\subsection{Enhanced insights from tidal asteroseismology.}
The sensitivity of the orbital evolution rates to perturbations in the stellar and orbital parameters implies in return access to tight observational constraint on these parameters, also known as tidal asteroseismology\cite{guo2022new}. 
The orbital evolution rates across the HR diagram display no discernible pattern (Fig. 4a), but this complexity is simply due to the superposition of $\dot{e}$ from every mode.
In contrast, each individual mode is coherent across the HR diagram (see Methods; Extended Data Fig. 1, 2).
Unfortunately, only the net orbital evolution rate is observationally accessible, but this suggests that measurements of individual mode frequencies and amplitudes can complement the integrated orbital evolution rates to provide a strong constraint on stellar structure and orbital configuration.

To highlight this further, we calculate the stellar flux variations due to tidally excited low-frequency g-modes\cite{townsend2003semi}.
We account for band-limitation of the observer and assume an edge-on viewing geometry.
Fig. 5a shows the stellar flux evolution over 400 Myr, where colored modes are those that exceed amplitudes of 10~p.p.m., which we take to be the limit of photometric observability.
The amplitudes of these tidally excited g-modes generally increase with decreasing $k$ due to the concentration of tidal forcing strength at low frequencies.
When the tidal forcing sweeps through resonance, modes can be excited to high amplitudes (10\%), with sustained excitation of modes up to $k=43$ (the present model only accounts for k$\leq$50) to amplitudes of 100~p.p.m. during resonance locking (Fig. 5a), consistent with Ref.\cite{de2017planet}.

Since high-frequency modes are often excited to amplitudes an order of magnitude above the observational limit, wavelength decomposition of the stellar flux can provide additional information about the nature of the excited mode and thus can be used as an additional measurement for probing the stellar interior.
Fig. 5b,d,e show the pulsation spectra in six independent wavelength bands between $0.6-5 \mu$m at three different stages in the pulsation evolution, with Fig. 5c,e,f showing the corresponding synthetic lightcurves.
Fig. 5b,c shows excitation of a single high-frequency mode with weak wavelength dependence.
Fig. 5d,e shows excitation of a single low-frequency mode with strong wavelength dependence.
This is useful since other sources of power at low orbital harmonics exist, including Doppler beaming and reflection of light off of the planet\cite{penoyre2019higher}.
These models of wavelength-dependent photometry can be used not only to probe the stellar interior, but also to refine models of these alternative mechanisms.
Fig. 5f,g shows excitation of two high frequency modes at once, each with a strong wavelength dependence.
This situation is likely less observationally prevalent, but is nonetheless important to understand since a resonance lock can be reinforced or broken by excitation of a second mode.
Understanding the dynamics of competing resonance locks thus depends on detailed characterizations of the photometric observability of multi-mode excitation.

In addition to providing insights into the stellar structure\cite{guo2022new}, the detection of high-frequency pulsations can also be observables for eccentric companions regardless of their orbital configuration. 
Such a strategy\cite{de2017planet} would be uniquely suited to eccentric and inclined systems\cite{millholland2019obliquity, alexander2023orbital}, whose planetary candidates may later be confirmed via traditional techniques such as radial velocity, astrometry, and/or direct imaging.

\subsection{Implications for tidal modification of stellar evolution.} 
Orbital angular momentum is usually assumed to be deposited in the star in a way that maintains rigid-body rotation. 
In reality, however, the deposition will be concentrated where the tidal energy is dissipated--typically, by the surface layers. 
Over timescales short compared to stellar evolution, this can establish a shear layer between the surface and the deeper envelope, as shown for self-excited oscillations\cite{townsend2018angular}. This can have two important consequences. First, the shear layer can trigger mixing via the Kelvin-Helmholtz instability. 
Second, a rich new phenomenology of ``differential resonances'' will be induced, through which only part of a star participates in a tidal resonance. 
It is also possible that angular momentum deposition in the chemically inhomogeneous zone just outside the boundary of the convective core\cite{townsend2018angular} can lead to mixing of the zone, which can grow the core size. 
The mass of the compact object is directly linked to the core size during the main sequence and later phases of evolution, so enhanced mixing has important implications for the subsequent evolution and fate of the star. 
We note that the only 3 known pulsating exoplanetary hosts\cite{de2017planet,kalman2022gravity,kalman2023discovery} are found at the intersection of the instability strip and the end of the main sequence, which we may warrant future investigations with a focus on linear and non-linear instabilities (i.e., possible triggering of self-excitation through a resonance).

\subsection{On the migration timescales and populations of hot Jupiters.}
Resonance locking may help explain the lack of planets currently observed to be tidally migrating due to both the enhanced circularization and migration rates leaving few planets observed in the process of migrating, and the increased migration efficiency towards the end of the main sequence resulting in young gas giants with mild migration rates and latent tidal migration potential (``dormant migrating gas giants''). 
The latter can be seen in the the skew of high orbital evolution rates toward the red-edge of the main sequence (Fig. 4a) and more rapid orbital evolution for older stars (Fig. 4b).
Tidal migration rates increase with stellar age due to the development of a g-mode cavity at the core-envelope boundary that can strongly couple to the tidal forcing (Fig. 1c). 
This leads to an increase in the number of resonances encountered and thus enhanced orbital migration rates.
The increase in stellar radius as the star ages also acts to increase the strength of the tidal forcing, accounting for an increase of $\sim 4\times$ over the main sequence lifetime of a $M=1.35M_\odot$ star.


Efficient tidal migration near the end of the main sequence points toward a tighter limit on the survival of close-in massive planets beyond the main sequence, with engulfment such as the one recently observed as the ultimate fate\cite{de2023engulf}.
This also suggests the possibility of two populations of migrating gas giants. 
A first population yields the hot Jupiters seen around early main-sequence stars which are unlikely to have formed due to resonance locking as introduced here. 
A second population evolves from ``dormant migrating giants'' on slowly-evolving eccentric orbits yielding hot Jupiters once their hosts start evolving off the main-sequence. 
Such hosts will often display high-amplitude tidally induced oscillations and support enhanced tidal migration rates.

\vspace{0.5cm}

\newpage
\pagebreak
\clearpage

\section{Methods}

\begin{methods}
\subsection{Stellar Evolution.}
We use the MESA stellar evolution code\cite{paxton2010modules,paxton2013modules,paxton2015modules,paxton2018modules,paxton2019modules,jermyn2023modules} to simulate the main sequence evolution of stars with masses in the range of $M=1.2 M_{\odot}$ to $M=1.5 M_{\odot}$, and metallicities in the range $Z=0.01-0.04$.
Stars of this type have a convective core, preventing excited g-modes from geometrically focusing in the core of the star, which can lead to nonlinearities from wave breaking\cite{weinberg2012nonlinear} for some massive exoplanets \cite{ma2021orbital}.
We calculate the linear tidal response with GYRE-tides which excludes inherently nonlinear dissipation mechanisms.
Models start at the zero-age main sequence and are evolved to the terminal-age main sequence, but the HR diagrams are truncated at the red-edge of the main sequence to avoid overlap with the Henyey hook.
We evaluate convective stability with the Ledoux criterion, though semiconvection and thermohaline mixing are found to have a negligible effect on the stellar evolution.
We include convective overshoot in the exponential scheme.
We do not consider mass loss or stellar wind since these are negligible in the type of main sequence stars considered in this study.
We neglect spin in the stellar evolution calculations because their evolution does not significantly change their spin rate, and magnetic braking is negligible for stars with a radiative envelope.
We do, however, consider the evolution of stellar spin in the orbital evolution calculations.
Example MESA inlists are provided in the supplementary materials.

\subsection{Tidal Asteroseismology for Orbital Evolution.}
We consider a binary system consisting of a star with mass $M_1$ and radius $R$, along with a point-mass companion of mass $M_2$.
The host star spins with a frequency $\Omega_\mathrm{rot}$.
We neglect Coriolis and centrifugal forces because low-mass main-sequence stars rotate well below their critical rotation rates, which means that that inertial waves cannot be resonantly excited.
We adopt a non-rotating reference frame and assume spin-orbit alignment.
Although spin-orbit misalignment is possible to include in the linear framework of GYRE-tides, its implementation in these calculations is beyond the scope of this work.
The companion orbits with a frequency $\Omega_\mathrm{orb}$ given by Kepler's third law
\begin{equation}
    GM_1(1+q)=a^3 \Omega_\mathrm{orb}^2,
\end{equation}
where $G$ is the gravitational constant, $q=M_2/M_1$ is the mass-ratio of the system, and $a$ is the orbital semi-major axis.
We study the tides raised on the host star due to the tidal portion of the secondary's gravitational potential.
Using a multipolar expansion in space and Fourier-series expansion in time, we express the tide-generating potential at position $\mathbf{r}$ and time $t$ as a superposition of partial tidal potentials
\begin{equation}
    \begin{aligned}
    \Phi_T(\mathbf{r},t) =& -\varepsilon_T \frac{GM}{R} \sum_{l=2}^{\infty} \sum_{m=-l}^{l} \sum_{k=-\infty}^{\infty} \bar{c}_{l,m,k} \left( \frac{r}{R} \right)^l \times\\
    & Y_{l}^{m}(\theta,\phi)exp\left[-ik\Omega_\mathrm{orb}(t-t_0) \right].
    \end{aligned}
\end{equation}
where $Y_{l}^{m}$ is a spherical harmonic of degree $l$ and order $m$, $t_0$ is the time of periastron passage, $\bar{c}_{l,m,k}$ is a tidal expansion coefficient given by Sun et al. (2023), and $\varepsilon_T$ is a dimensionless parameter describing the strength of the tidal forcing given by
\begin{equation}
    \varepsilon_T = \left( \frac{R}{a} \right)^3 \left( \frac{M_2}{M_1} \right).
\end{equation}
For hot Jupiters orbiting intermediate-mass main-sequence stars, $\varepsilon_T \sim 10^{-5}$. 

The tide-generating potential perturbs the spherical symmetry of the star's gravitational potential, which in turn perturbs the orbit of the companion from a pure Keplerian orbit.
The magnitude and direction of these perturbations to the orbital parameters depend on how dissipation of the tidally-induced pulsations acts to transfer energy and angular momentum between the star and the orbit.
The star's tidal response consists of the equilibrium and dynamical tides.
We account for damping of the equilibrium tide by turbulent viscosity in the convective envelope, as well as by radiative damping of dynamically excited internal gravity waves (the dynamical tide). 
We use the GYRE-tides stellar pulsation code\cite{sun2023tides} to directly solve the linear, non-radial, non-adiabatic stellar oscillation equations with an inhomogeneous tidal forcing term.
Example GYRE-tides inlists are provided in the supplementary materials.

The changes in orbital elements through these mechanisms are given by
\begin{equation}\label{eq:orbev_dedt}
    \begin{aligned}
    \left( \frac{de}{dt} \right)_{sec} = 4 \Omega_\mathrm{orb} q \sum_{l,m,k\geq 0} &\left( \frac{R}{a} \right)^{l+3} \left( \frac{r_s}{R} \right)^{l+1} \\
    &\times \kappa_{l,m,k} \text{Im} (\Bar{F}_{l,m,k}) \Bar{G}^{(3)}_{l,m,k}(e)
    \end{aligned}
\end{equation}
where the $sec$ subscript denotes a secular (orbit-averaged) quantity.
$\Bar{F}_{l,m,k}$ are dimensionless quantities that measure the response of the star to the various forcing frequencies $\sigma_{m,k}$.
$\kappa_{lmk}$ are the same as those given by Willems et al. (2010). 
$\bar{G}^{(3)}_{lmk}$ is given by Sun et al. (2023), and differs only by sign convention from the coefficients derived by Smeyers et al. (1998) and Willems et al. (2003).
Similar $\bar{G}$ coefficients exist for changes in the semi-major axis, stellar rotational angular momentum, and argument of periastron.
We account for the effect of rotation only in its Doppler-shift of the co-rotating frame forcing frequency.

While eq. 2 and eq. 4 provide the exact expressions for the orbital evolution rate, computational considerations require us to truncate the summations in $l$ and $k$. 
Since the orbital evolution rate scales with $(R/a)^{l+3}$, where $(R/a) \approx 0.1$, the strength of tidal forcing falls off rapidly above $l=2$.
We restrict $m$ by noting that $l=2$, $m=\pm 1$ modes are not excited by the tidal potential, and we find that orbital evolution rates due to $l=2$, $m=-2$ modes are consistently several orders of magnitude smaller than for $m=0,2$ modes.
We calculate only the $l=2$, $m=0,2$ modes\cite{witte2002orbital}.
The orbital evolution rates also decrease for sufficiently high $k$, and we truncate the summation at $k=50$. 

\subsection{Benchmarking accelerated tidal migration.}
In order to evaluate whether these tidal processes can contribute to rapid circularization and inward migration, we compare the orbital evolution produced by GYRE-tides and by a constant-Q model of stellar and planetary tides from Jackson et al. (2008).
In this framework\cite{goldreich1966q,jackson2008tidal}, the rate-change in eccentricity is given by
\begin{equation}
    \frac{de}{dt}=-e \left[ \frac{63}{4}(G M_1^3)^{1/2} \frac{R_2^5}{Q_p M_2} + \frac{171}{16} (G/M_1)^{1/2} \frac{R_1^5 M_2}{Q_\ast} \right]a^{-13/2},
\end{equation}  
where $R_1$ is the stellar radius, $R_2$ is the planetary radius, and  $Q_\ast$ and $Q_p$ are quality factors for the star and the planet, respectively.
A similar expression exists for $da/dt$.
Jackson et al. (2008) estimate $Q_\ast=10^{5.5}$ and $Q_p=10^{6.5}$, though they consider a wide range of possible values from $10^4-10^8$ in each parameter.
Extended Data Fig. 4 compares the variability in tidal evolution trajectories for the GYRE-tides model to the constant-Q model.
Parameters in the constant-Q model are perturbed from $-3 \sigma$ to $+3 \sigma$ in each parameter, while the perturbations to the GYRE-tides model are given in the legend to the right of Extended Data Fig. 4. 
While the perturbations to parameters in the GYRE-tides model are generally smaller (e.g. perturbations of $1 \sigma$ in $M_\odot$ in the GYRE-tides model vs. perturbations of $3 \sigma$ in the constant-Q model), the resulting variability in tidal evolution trajectories is much higher for the GYRE-tides model.
This is especially true for well-constrained parameters such as $a$, where no deviation of the tidal evolution trajectories is visible in the constant-Q case, while significant divergence in tidal trajectories emerges after only $\sim 50$ Myr for the GYRE-tides model. 
Variations in $e$ in the constant-Q model result in parallel tidal evolution trajectories that maintain the initial uncertainty in $e$. 
The GYRE-tides model, on the other hand, increases the difference between tidal evolution trajectories as time elapses. 
Of particular interest are the variations in the stellar mass.
The constant-Q model vastly underestimates the variation in tidal evolution trajectories due to variations in stellar mass. 
Perturbations of $\pm 3 \sigma$ in stellar mass result in smaller variations than perturbations of $\pm 1 \sigma$ in the GYRE-tides model. 

We also see that the tidal evolution trajectories are most sensitive to the choice of stellar quality factor, with the planetary quality factor playing a less significant role. 
The wide range of orbital evolution rates resulting from varying $Q_\ast$ over four orders of magnitude provides a convenient way to benchmark whether a tidal evolution trajectory displays accelerated tidal migration relative to the baseline $Q_\ast=10^{5.5}$. 
$Q_\ast=10^{5.5}$ results in $\dot{e}$ on the order of $10^{-10}$/yr over the studied period.

The background $\dot{e}$ is on the order of $10^{-11}$/yr for the GYRE-tides model, so without resonance the circularization timescale should be an order of magnitude larger (Extended Data Fig. 3, Supplementary Figs. 1, 2). 
However, brief resonances sweep through $\dot{e}\sim 10^{-5}$/yr and resonance locks sustain rates of $10^{-8}$/yr, which compensates for the lower background orbital evolution rates.
In cases where few resonances are encountered, such as for young main sequence stars ($t_0 - 1\sigma$), the equivalent $Q_\ast$ is as high as $\sim10^7$. 
For cases where many resonances are encountered, such as for old main sequence stars ($t_0 + 1\sigma$), the equivalent $Q_\ast$ is closer to $\sim 10^{5}$.
Similarly, perturbing the stellar mass by $+1 \sigma$ results in an effective $Q_\ast$ of $10^{4.5}$.
Increasing the mass of the star while holding the age constant has the same effect as advancing the age while holding the mass constant since higher mass stars have short main sequence lifetimes.
This emphasizes that tides can be an effective driver of accelerated orbital evolution in old main sequence stars, but are unlikely to accelerate orbital migration in young main sequence stars.

\subsection{Limits of orbital trajectory predictability.}
The deviation of orbital evolution trajectories due to small perturbations in stellar properties is of fundamental importance to understanding whether tidal evolution calculations can be used to deterministically predict the future state of exoplanet systems or reconstruct their formation state. 
We characterize the predictability of tidal evolution trajectories through a series of fixed-orbit calculations similar to those in fig. 4a, but we seek to understand whether features of the net orbital evolution rate are preserved under small perturbations to the stellar parameters. 
We simulate the stellar evolution of 15 stars, with perturbations in mass and metallicity ranging from 0.1 to 5\% with the baseline star from the main text ($M=1.36 M_\odot$, $Z=0.02262$) used as a starting point.
For each of these stars, we calculate the orbital evolution rates for a fixed-orbit tidal perturber for 8 different spin-orbit states; for similarity to fig. 4, we focus on an orbit with $e=0.575$, $a=0.153$ au, and a pseudosynchronous stellar spin rate. 
Individual resonances can be easily matched between the reference star and those with small mass perturbations (say, 0.1\%) (Extended Data Fig. 5a-d).
Larger mass perturbations distort the shape and location of these resonances, and only the general increase in density of resonances toward the end of the main sequence remains visible at perturbations of $5\%$.
Perturbations in metallicity result in smaller deviations in the shape of orbital evolution rates, though the location of individual resonances in time is nonlinear with increases in $Z$, another representation of the zig-zag pattern in Extended Data Fig. 2b. 

We now calculate how this variation in orbital evolution rates affects predictability of the orbital migration trajectories; we quantify the predictability of an orbital trajectory via the timescale over which its deviation from a reference trajectory grows to observable levels (say, $1\sigma$ in e or a). 
We estimate this timescale by calculating the instantaneous deviation in the orbital evolution rates. 
For eccentricity, this is 
\begin{equation}
    \Delta\dot{e}= \dot{e}(M_i, Z_i, t) - \dot{e}(M_0, Z_0, t),
\end{equation}
where $M_0$, $Z_0$ are the reference mass and metallicity and $M_i$, $Z_i$ are the perturbed mass and metallicity, $t$ is a common stellar age. 
We define a timescale over which the orbital trajectories remain observationally indistinguishable, 
\begin{equation}
    T_e = \frac{\sigma_e}{\Delta \dot{e}}.
\end{equation}
To make this calculation in practice, we begin with time series of $\dot{e}$ for a fixed spin-orbit configuration for a range of stellar mass and metallicity perturbations (Extended Data Fig. 5, Extended Data Fig. 6). 
We aggregate these $\dot{e}$ values into a distribution (Extended Data Fig. 5q-r, Extended Data Fig. 6q-r), revealing that the frequency-rate distribution is distributed according to a power law with slope $-1/2$. 
This long-tailed distribution of orbital evolution rates indicates the importance of rare episodes of high orbital evolution rate in controlling the orbital migration even before considering resonance locking.
The other implication of this distribution is that there is no well-defined average orbital evolution rate.

We also aggregate the $\Delta\dot{e}$ values for each perturbed stellar model into a distribution (Extended Data Fig. 5s-t, Extended Data Fig. 6s-t). 
The deviation in orbital evolution rates, $\Delta\dot{e}$, grows with mass perturbations, but is more stable for metallicity perturbtaions (Extended Data Fig. 6c).
To estimate $T_e$, we sample the distribution of $\Delta\dot{e}$ for orbital evolution rates, which we integrate to produce an ensemble of orbital trajectories.
For each member of the ensemble, we record the time needed to accumulate a change of at least $\sigma_e=0.011$, the uncertainty in $e$ for HAT-P-2b as reported in Ref.\cite{de2023revisiting}.
These trajectories do not possess the same intermittent migration structure due to the removal of time-dependence when constructing the distribution, but the ensemble allows us to estimate a distribution of $T_e$.
The distribution of $T_e$ shows that larger mass perturbations require less time to accumulate observable deviations in orbital evolution rates, i.e. larger mass differences accumulate differences faster (Extended Data Fig. 5u-v, Extended Data Fig. 6u-v). 
For the spin-orbit state and stars considered in this example, eccentricity differences become observable over $\sim5-15$ Myr timescales.

We also calculate the distribution of $\dot{e}$ and $\dot{a}$ for the live-orbit simulations with a coevolving star and orbit (Extended Data Fig. 7a-l). 
Coevolution of the stellar and orbital configuration modifies the distribution such that it has a well-defined peak. 
The live-orbit distribution is depleted in the lowest $\dot{e}$ and doesn’t reach the highest $\dot{e}$ achieved in the fixed-orbit simulation. 
We also calculated $T_e$ and $T_a$ for the live-orbit case (Extended Data Fig. 7m-x). 
We did this by calculating the time needed to accumulate a $1\sigma$ deviation in $e$ or $a$ from every point in the orbital trajectories. 
That is to say, we begin with the initial orbital configuration at $t_0$ and find the first $t$ such that $t>t_0$ and $e(t)-e(t_0)>\sigma_e$. 
We see that the coevolution of the star and orbit enables a much broader range of $T_e$, with the most likely $T_e$ around 1 Myr, though several rare cases reach as low as 1 kyr or as high as 1 Gyr. 
This indicates that models of tidal evolution and backpropagation are robust in comparison to observational constraints over timescales of typically 1 Myr.

\subsection{Tidal asteroseismology for photometric observations.}
We calculate the stellar flux variations due to tidally excited low-frequency g-modes\cite{townsend2003semi}.
Stellar flux variations are due to a combination of radial displacements at the stellar surface, and Lagrangian perturbations to the radiative luminosity\cite{burkart2012tidal,fuller2012dynamical}.
Rather than assuming black body radiation, we use the CAP18 photospheric grids\cite{prieto2018collection} to estimate the stellar flux variations between wavelengths of $0.6-5.0 \mu$m at each spherical and orbital harmonic.
Since these photospheric models are not calculated for stars of precisely the mass, metallicity, age, etc. that we study, we use the MSG code\cite{townsend2023msg} to interpolate the stellar spectra.
Finally, we calculate the observable flux variations by averaging the luminosity variations over the visible disk.
We assume  an edge-on viewing geometry, but do not include transit effects.

Figure 4a shows the evolution of the observed flux variation over the full $0.6-5.0 \mu$m wavelength range.
The tidal forcing is strongest for low frequencies, leading to persistent power on the $k<13$ orbital harmonics at the $10^{-5}$ level.
During resonance between the tidal forcing and a stellar eigenmode, additional higher frequency modes can be excited to observable levels (Fig. 5b,c), low-frequency modes can be excited to higher amplitudes (Fig. 5d,e), and multiple modes can be excited at once (Fig. 5f, g).
Also visible in Fig. 5a is complex resonance locking dyanmics.
For example, excitation of the $k=32$ mode is sustained between $2586-2588$ Myr at an amplitude of $\sim 10^{-4}$ (Fig. 5a,b,c). 
$k = 4, 14, 16, 30$ modes are resonantly excited during this time, but the resulting orbital evolution is not enough to break the resonance lock with the $k = 32$ mode. 
Excitation of the $k = 22$ mode at $2588$ Myr eventually breaks the resonance lock, and the mode amplitude oscillates over the next $8$ Myr until another resonance lock with the $k = 30$ mode is established.
Although individual modes are typically locked into resonance for $<5$ Myr, multiple resonance locks can be chained together resulting in sustained photometric observability over timescales of $\sim50$ Myr.

In addition to structure in the dynamics of resonance locking through time, the tidal response of the star has interesting structure across the HR diagram.
For example, although the net orbital evolution rate has no clear pattern across the HR diagram (Fig. 4a), this is simply due to the superposition of many different modes, each with a coherent and unique fingerprint across the HR diagram (Extended Data Fig. 1). 
The orbital evolution due to a particular mode is not observable since the orbit changes due to a summation over all modes.
However, photometric observations enable mode identification, and individual modes also have a coherent photometric signature across the HR diagram (Extended Data Fig. 2). 
This tidal asteroseismology can thus be used to place strong constraints on stellar structure.
For example, if both a $k=10$ and a $k=30$ mode were identified (Extended Data Fig. 2), the space of stellar structures can be strongly constrained to a few specific points on the HR diagram.
That is to say, multiple mode excitation constrains the possible stellar model to the intersection of the sets of resonances in Extended Data Fig. 2a and 2b. 

\subsection{Numerical integration for orbital trajectories.}
In order to convert the estimates of orbital evolution rates (e.g. $\dot{e}$) into trajectories in the orbital parameters, we solve a system of differential equations
\begin{equation}
    \frac{dy^m_n}{dt} = f_n(t,y)
\end{equation}
iteratively for the spin-orbit state
\begin{equation}
    y^m_n(t+dt_n)
\end{equation}
where $dt_n$ is the $n^{th}$ timestep and $y^m_n$ stands in for the $m^{th}$ spin-orbit parameter (eccentricity, semi-major axis, or stellar rotation rate).
We update the orbital elements and stellar spin frequency at each time step with the Runge-Kutta-Fehlberg method\cite{press1992adaptive}, a fourth-order accurate method embedded in a fifth-order method used for error estimation.
Error estimation allows for adaptive timestepping depending on the current dynamics of the system, i.e. the timestep is reduced as the tidal forcing sweeps through resonances with the stellar eigenmodes.
With this method, the solution is a weighted average of six increments to the spin-orbit parameters
\begin{equation}
    y^m_{n+1} = y^m_n + \sum_{i=1}^{6} c_{i} k_{i} + \mathcal{O}({dt^6}).
\end{equation}
where the intermediate increments are given by
\begin{equation}
    k_i = dt_n \frac{dy^m}{dt} \left( t_n + a_i dt_n; y^m_n + \sum_{j<i} b_{ij} k_i \right). \\
\end{equation}
The embedded fourth-order formula gives an alternate estimate of the updated system state
\begin{equation}
    y^{m \ast}_{n+1} = y^m_n + \sum_{i=1}^{6} c^{\ast}_{i} k_{i} + \mathcal{O}({dt^5}).
\end{equation}
The parameters $a_i$, $b_{ij}$, $c_i$, and $c_{i}^{\ast}$ can be found in a Butcher tableau for the Runge-Kutta-Fehlberg method\cite{press1992adaptive}.
The error estimate is given by
\begin{equation}
    \Delta^{m}_{n+1} = y^m_{n+1}-y^{m \ast}_{n+1} = \sum_{i=1}^{6} (c_i - c^{\ast}_i) k_i.
\end{equation}
For each orbital element, the error between the 4th and 5th order solutions suggests a new timestep, given by
\begin{equation}
    dt^{m}_{n+1} = 
    \begin{cases}
        S \cdot dt_n\left( \frac{\Delta_{max}}{\Delta^{m}_{n+1}} \right)^{1/5} &  (\Delta_{max} \geq \Delta^{m}_{n+1}) \\
        S \cdot dt_n\left( \frac{\Delta_{max}}{\Delta^{m}_{n+1}} \right)^{1/4} &  (\Delta_{max}\leq \Delta^{m}_{n+1})
    \end{cases}
\end{equation}
where $S$ is a safety factor we set to $0.9$.
The new timestep is chosen to be the most conservative of the proposed timesteps
\begin{equation}
    dt_{n+1} = min \{ dt_{n+1}^{m}, dt_{max} \}_m
\end{equation}
where $dt_{max}$ is an additional safeguard to ensure we do not miss brief resonances due to extending time steps between resonances.
Timesteps near $dt_{max}$ generally indicate slow orbital evolution, and this parameter is empirically set to $dt=100$ kyr in order to avoid aliasing the stellar evolution.
Smaller timesteps indicate rapidly changing orbital evolution rates, where the timestep is reduced to maintain errors in the orbital parameters below some fractional error threshold; we set this threshold to $10^{-9}$, i.e. the update in accurate to one part in a billion.

\subsection{Stellar model interpolation.}
Full coupling of stellar and orbital evolution requires the ability to solve for the tidal response of a stellar model at arbitrary times.
This means that we either need to fully couple the MESA and GYRE-tides codes, or else devise some interpolation scheme whereby the tidal response of the interpolated stellar model closely matches that of the stellar model produced by the stellar evolution code.
We prefer the second approach since it permits decoupling of the stellar evolution and orbital evolution segments of the calculation.
This interpolation scheme must adapt to changes in radius and radial sampling of the star while preserving spatial resolution around key features of the stellar interior such as the boundary between the convective core and radiative envelope and near the stellar surface.

We adopt a simple linear interpolation between pairs of points in two stellar models.
This requires a sufficiently small timestep in the stellar model such that perturbations in stellar structure are locally linear; we use a maximum timestep of $10$ kyr in the stellar models.
Interpolation errors can introduce false peaks in the orbital evolution rates, leading to misinterpretation as being due to resonances (Supplementary Fig. 3a,b,c).
However, interpolation errors do not become significant until $dt>200$ kyr (Supplementary Fig. 3d,e,f), with timesteps up to $1$ Myr retaining the same general shape of the orbital evolution rates (Supplementary Fig. 3a,b,c).

When the number of radial samples is equal between stellar models, the interpolation is trivial since the radius provides a natural ordering and thus natural coupling.
When the number of radial samples differs, we use an optimal transport based method for first finding the optimal matching between points between the two curves, and then perform the interpolation with the coupled points\cite{chewi2021fast, bryan2023capturing}.
To apply optimal transport, we reinterpret the cumulative number of points as a function of radius as a cumulative distribution function.
The optimal transport map between the two radial sample profiles is then the map which matches quantiles.
This allows single points to match with multiple other points (i.e. allows mass splitting), which accounts for the difference in sampling.
We approximate one stellar model at the sampling of the other by barycentric projection of the optimal transport map.
This produces radial profiles with the same sampling, admitting a natural pairing of points, and thus allows simple linear interpolation to produce a stellar model at any arbitrary time.

\newpage
\pagebreak
\clearpage

\begin{addendum}
\item[Data Availability]
We make use of the orbital configuration of HAT-P-2b and the stellar properties of HAT-P-2\cite{bonomo2017gaps,stassun2019revised}.
The stellar tidal response was converted to synthetic photometry using the CAP18 photospheric grids\cite{prieto2018collection}.
Simulation results are available on Zenodo\cite{bryan2024zenodo}. 

\item[Code Availability] This work makes use of the following publicly available codes: \texttt{MESA}\cite{paxton2010modules,paxton2013modules,paxton2015modules,paxton2018modules,paxton2019modules,jermyn2023modules} for stellar evolution, \texttt{GYRE}\cite{townsend2013gyre,townsend2018angular,goldstein2020contour,sun2023tides} for calculating stellar pulsation properties and orbital evolution rates, and \texttt{MSG}\cite{townsend2023msg} for converting the stellar tidal response into synthetic photometric observables.
Tools for coupling these codes for live-planet simulations is publicly available at\\ \texttt{https://github.com/jaredbryan881/orbev}.

\item[Acknowledgements] J.B. and Z.L.d.B. acknowledge the National Science Foundation for supporting this work through the Graduate Research Fellowship program under Grant No. 1745302 and the MIT Presidential Fellowship.
The simulations presented in this paper were performed on the Engaging cluster at the MGHPCC facility.

\item[Author Contributions]
J.d.W. designed the study. J.B. developed the computational framework for the study with notable support from M.S. and R.H.D.T regarding the stellar and orbital modeling and from J.d.W. and Z.L.d.B. regarding exoplanetary context and observables. All authors contributed to the manuscript writing, which was led by J.B. 

\item[Competing Interests] The authors declare that they have no competing financial interests.

\item[Correspondence] Correspondence and requests for materials
should be addressed to J.B.~(email: jtbryan@mit.edu) or J.d.W. (email: jdewit@mit.edu). 

\end{addendum}

\newpage

\section{Figure Legends/Captions (for main text figures)}
\includegraphics[scale=1.5]{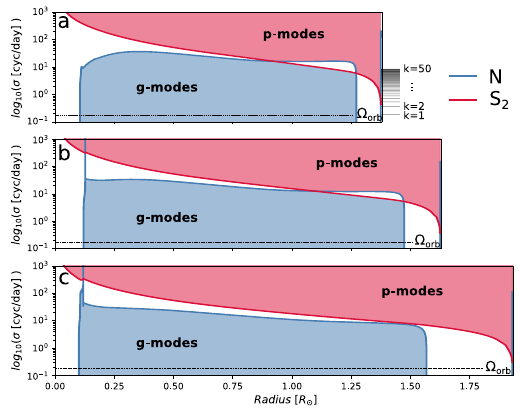}\\
\textbf{Figure 1. The rich tidal forcing spectrum of eccentric orbits.} Snapshots of the structure of an $M=1.36 M_\odot$, $Z=0.02262$ star as represented by a propagation diagram at three times: (a) Zero-age main sequence, (b) midway through the main sequence, and (c) Red-edge of the main sequence. The Blue line is the Brunt-V\"{a}is\"{a}l\"{a} frequency $N$ and the red line is the $l=2$ Lamb frequency $S_2$. g-modes can propagate in the blue region, where the pulsation frequency $\sigma$ is below both $N$ and $S_2$. The orbital frequency $\Omega_{orb}$ is plotted as a line near the bottom, with harmonics $0<k<50$ of the orbital frequency shown adjacent to (a). Circular orbits excite only the $k=2$ harmonic, while eccentric orbits spread energy across all $k$.

\includegraphics[scale=0.8]{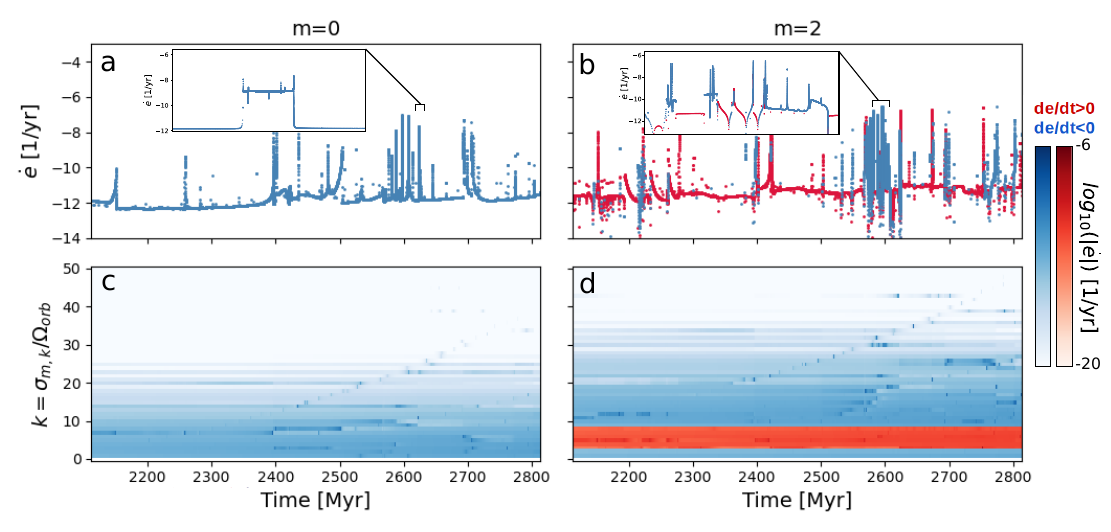}\\
\textbf{Figure 2. Episodic and wandering tidal migration by alternating resonance locks.} Rate-change of the orbital eccentricity induced by excitation of quadrupolar (l=2) modes with (a) $m=0$ and (b) $m=2$ by a tidal perturber with evolving orbital configuration. The host star has $M=1.36 M_\odot$ and $Z=0.02262$. The baseline $\Omega_\mathrm{rot}$ is set to 1.5$\Omega_\mathrm{ps}$, the crossover point for the tidal torque in Sun et al. (2023). Red and blue points correspond to positive and negative $de/dt$, respectively. Inset in (a) shows $5$ Myr of tidal evolution centered on a resonance lock with the $m=0$, $k=12$ mode, and the inset in (b) shows $25$ Myr of tidal evolution with an isolated resonance lock with the $m=2$, $k=30$ mode followed by alternating resonance locks with the $m=2$, $k=32$, $22$, $15$, $30$, and $19$ modes. The rate-change of the orbital eccentricity is decomposed into orbital harmonics $k=\sigma_{m,k}/\Omega_\mathrm{orb}$ for (c) $m=0$ and (d) m=2.

\includegraphics[scale=1.0]{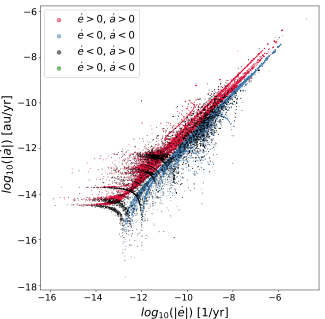}\\
\textbf{Figure 3. Diversity of senses of orbital evolution.} The magnitude of the rate-change of the semi-major axis and of the eccentricity corresponding to the star-planet configuration of Fig. 2. The color of the points corresponds to the sense of orbital evolution, with red indicating increases in both the semi-major axis and orbital eccentricity, blue indicating decreases in both, and black denoting circularization but expansion of the orbit.

\includegraphics[scale=1.5]{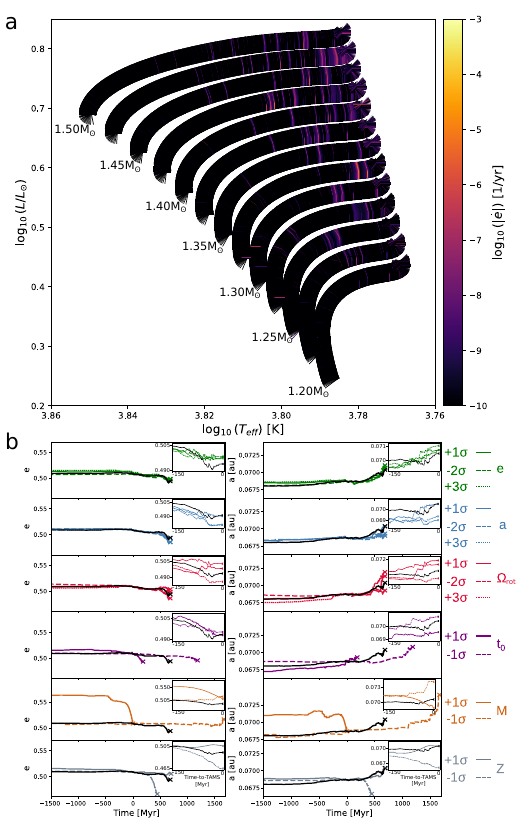}\\
\textbf{Figure 4. Sensitivity of orbital evolution to stellar and orbital initial conditions.} (a) Rate-change of eccentricity across the HR diagram from the zero-age main sequence to the red-edge of the main sequence for a fixed orbital configuration, corresponding to the initial orbital state of the black curve in Fig. 4b, for a fixed $Z=0.02$ and a range of masses. The colormap is truncated below at $\dot{e} \approx 10^{-10}/yr$ to emphasize resonances. (b) Forward and time-reversed orbital evolution trajectories with perturbations in (green) eccentricity $e$, (blue) semi-major axis $a$, (red) stellar rotation frequency $\Omega_{rot}$, (purple) initial stellar age $t_0$, (orange) stellar mass $M$, and (grey) stellar metallicity $Z$. Colors and perturbation sizes are given by the legend on the right. The ``x" ending each trajectory marks the terminal age main sequence. The insets in (b) show a zoom of tidal evolution over the final 150 Myr of the main sequence. 

\includegraphics[scale=0.75]{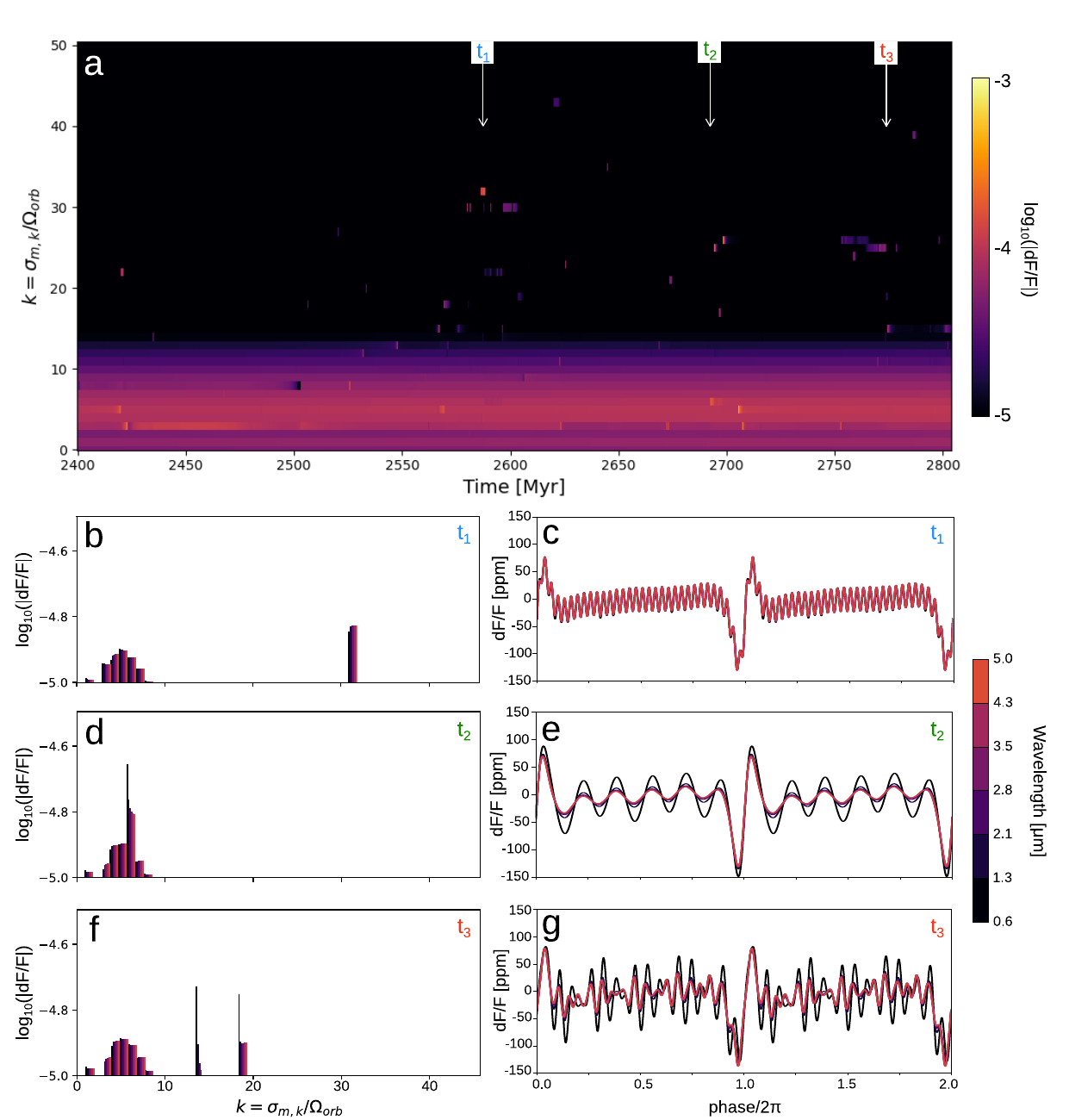}\\
\textbf{Figure 5. Observability of Tidally Excited Oscillations} (a) Evolution of the spectrum of relative flux variations over wavelengths of $0.6-5$ $\mu$m. (b,d,f) Stellar pulsation spectra at three times decomposed into $6$ wavelength bins between $0.6-5$ $\mu$m. Lines of different wavelength are all on integer $k$, but are offset for visibility. (c,e,g) Wavelength-dependent synthetic lightcurves corresponding to the pulsation spectra in (b,d,f). The pulsating star has $M=1.36 M_\odot$ and $Z=0.02262$.

\renewcommand{\figurename}{Supplementary\,figure}

\section{Tables and Figures for Methods}
\includegraphics[scale=0.1]{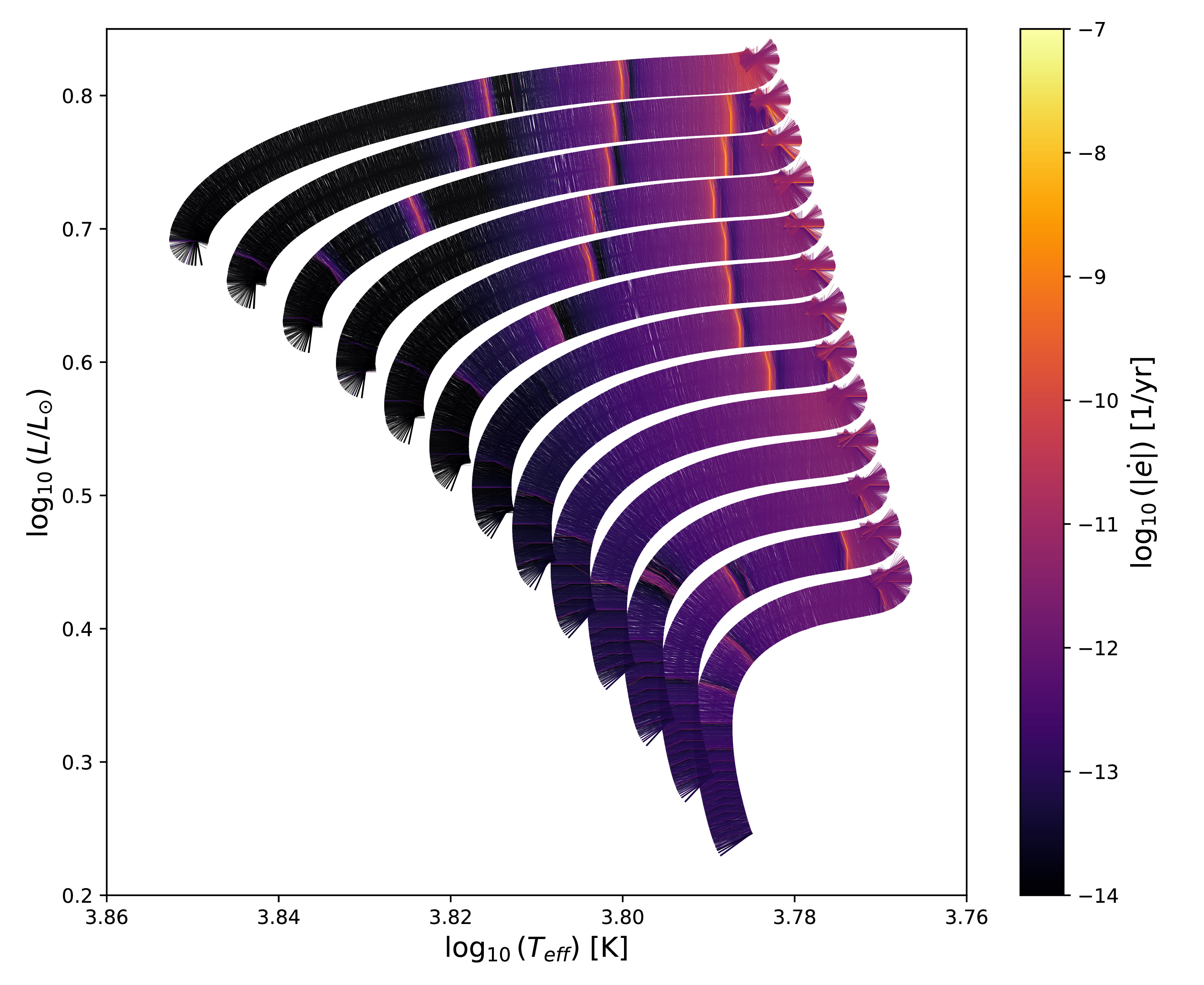}\\
\textbf{Extended Data Figure 1. Coherence of single mode damping rates across stellar models.} HR diagram showing the contribution to $de/dt$ of the $m=2$, $k=10$ orbital harmonic excited by tidal forcing with a fixed orbital configuration and a fixed $Z=0.02$ for a range of masses.

\includegraphics[scale=0.075]{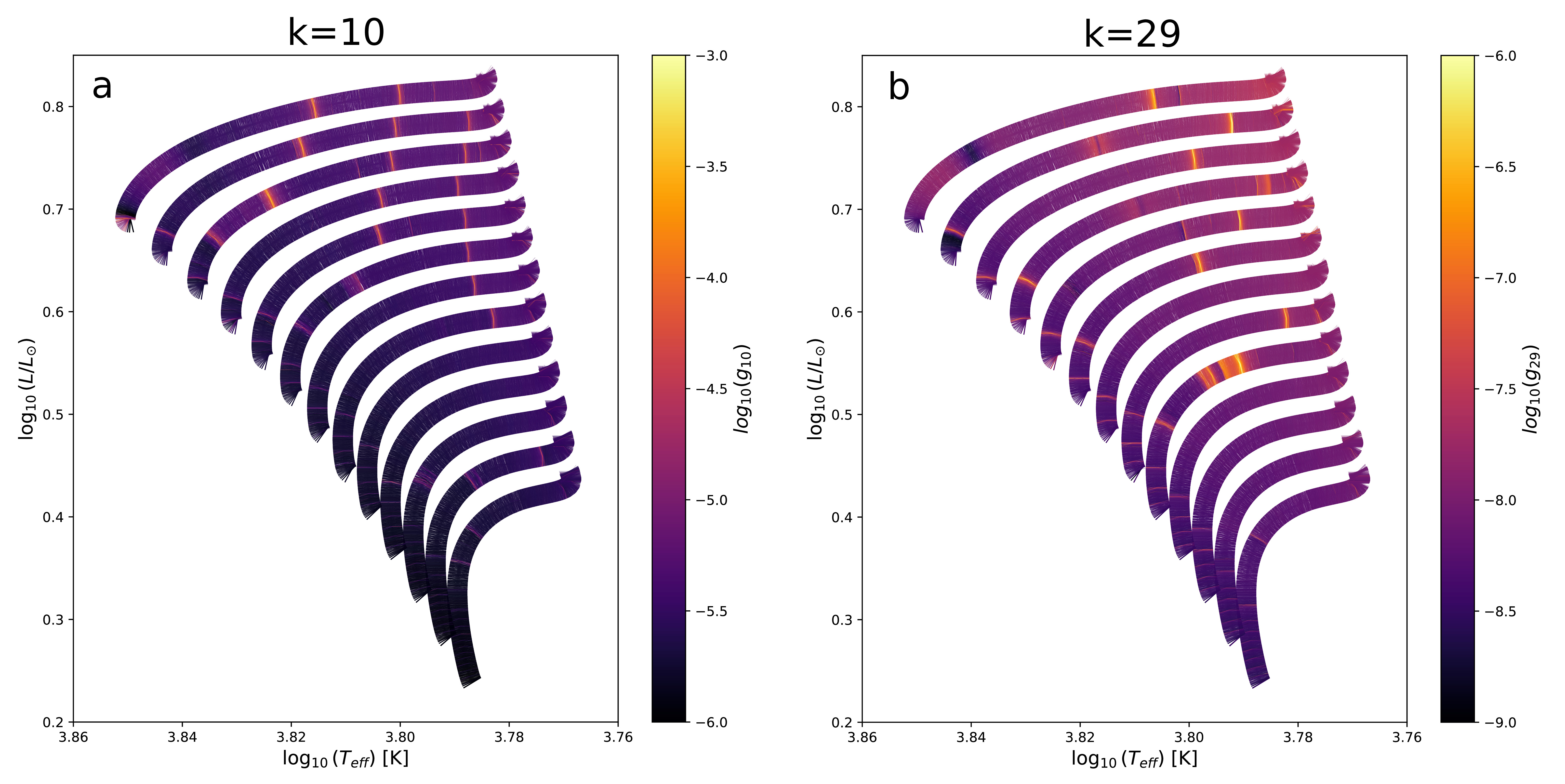}\\
\textbf{Extended Data Figure 2. Coherence of individual modes across stellar models.} HR diagrams of the amplitude of the (a) $k=10$ and (b) $k=30$ orbital harmonics excited by tidal forcing of a host star with $M=1.36 M_\odot$ and $Z=0.02262$ with a fixed orbital configuration. Amplitudes are for an edge-on view angle.

\includegraphics[scale=0.075]{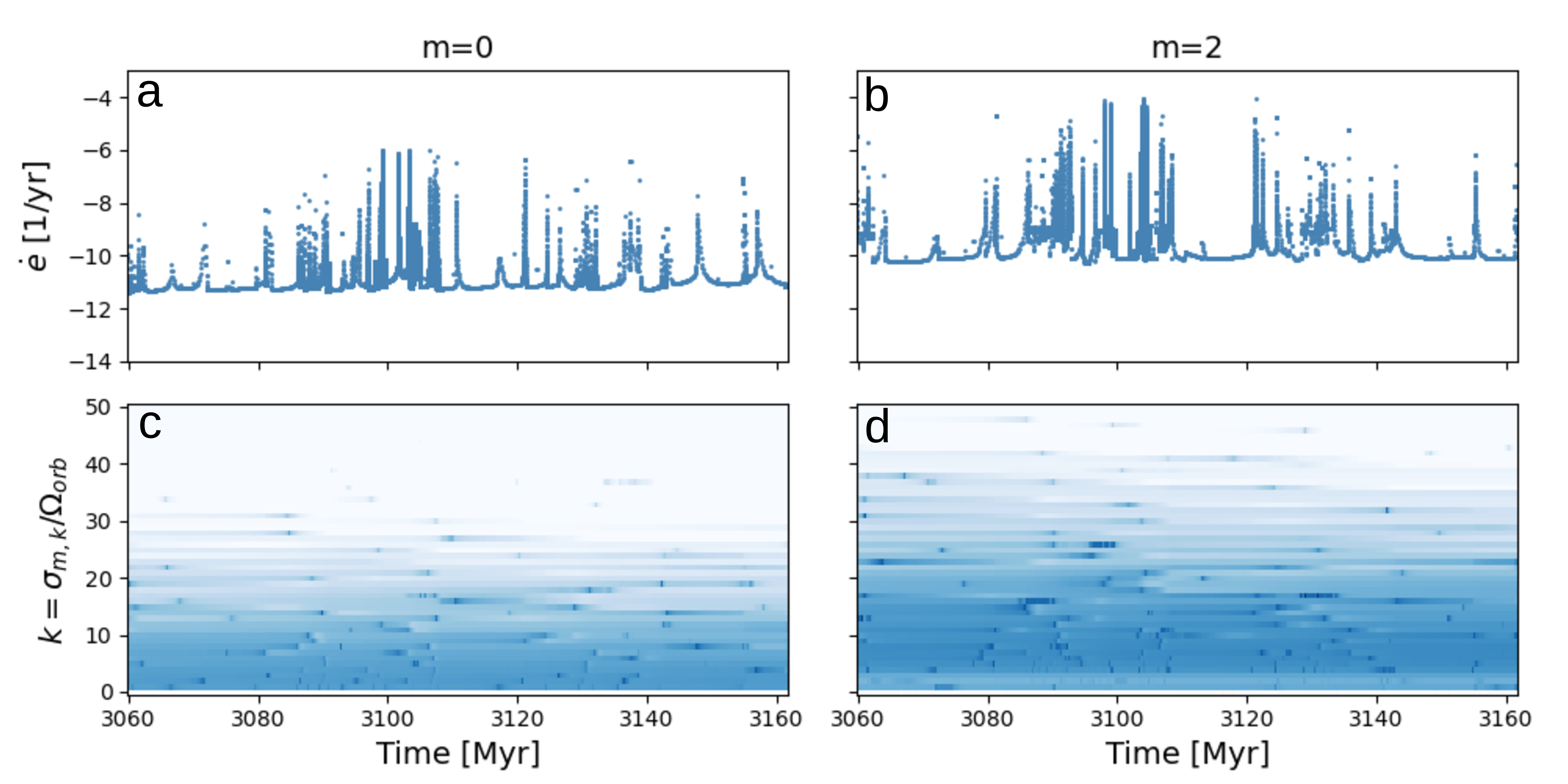}\\
\textbf{Extended Data Figure 3. Decomposition of tidal response of a slowly rotating star into different modes.} (a-b) Rate-change of the orbital eccentricity induced by excitation of quadrupolar (l=2) modes with $m=0,2$ in a host star with $M=1.36 M_\odot$ and $Z=0.02262$ by a tidal perturber with evolving orbital configuration. The tidal migration in this case begins at $t_0+1\sigma$. Red and blue points correspond to positive and negative de/dt, respectively. (c-d) Rate-change of the orbital eccentricity decomposed into orbital harmonics $k=\sigma_{m,k}/\Omega_\mathrm{orb}$. The baseline $\Omega_\mathrm{rot}$ is set to the value estimated for HAT-P-2 by Bonomo et al. (2017). Colors correspond to those in Fig. 2.

\includegraphics[scale=0.1]{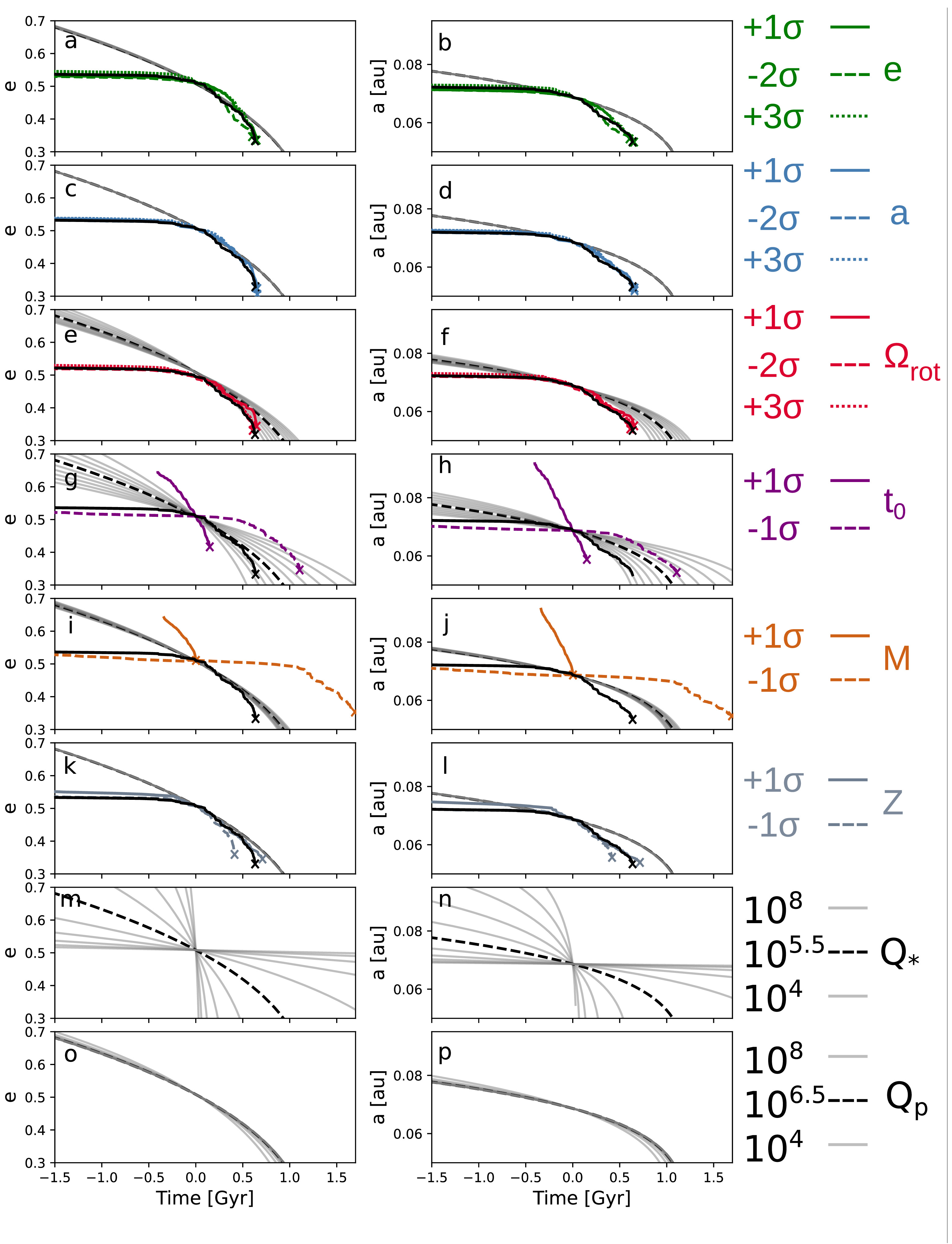}\\
\textbf{Extended Data Figure 4. Accelerated tidal evolution by intermittent resonance locking.} Comparison between GYRE-tides and constant-Q tidal evolution models. Forward and time-reversed orbital evolution trajectories with perturbations in each of the orbital and stellar parameters. The baseline $\Omega_\mathrm{rot}$ is set to the value estimated for HAT-P-2 by Ref.\cite{bonomo2017gaps}. The solid black line corresponds to the baseline GYRE-tides model. Colored lines correspond to those in Fig. 4b. The dotted black line corresponds to the best-fit constant-Q model from Jackson et al. (2008). Gray lines are evenly spaced between $\pm 3 \sigma$ in the given parameter except for $Q_\ast$ and $Q_p$, who span the full range considered by Jackson et al. (2008).

\includegraphics[scale=0.1]{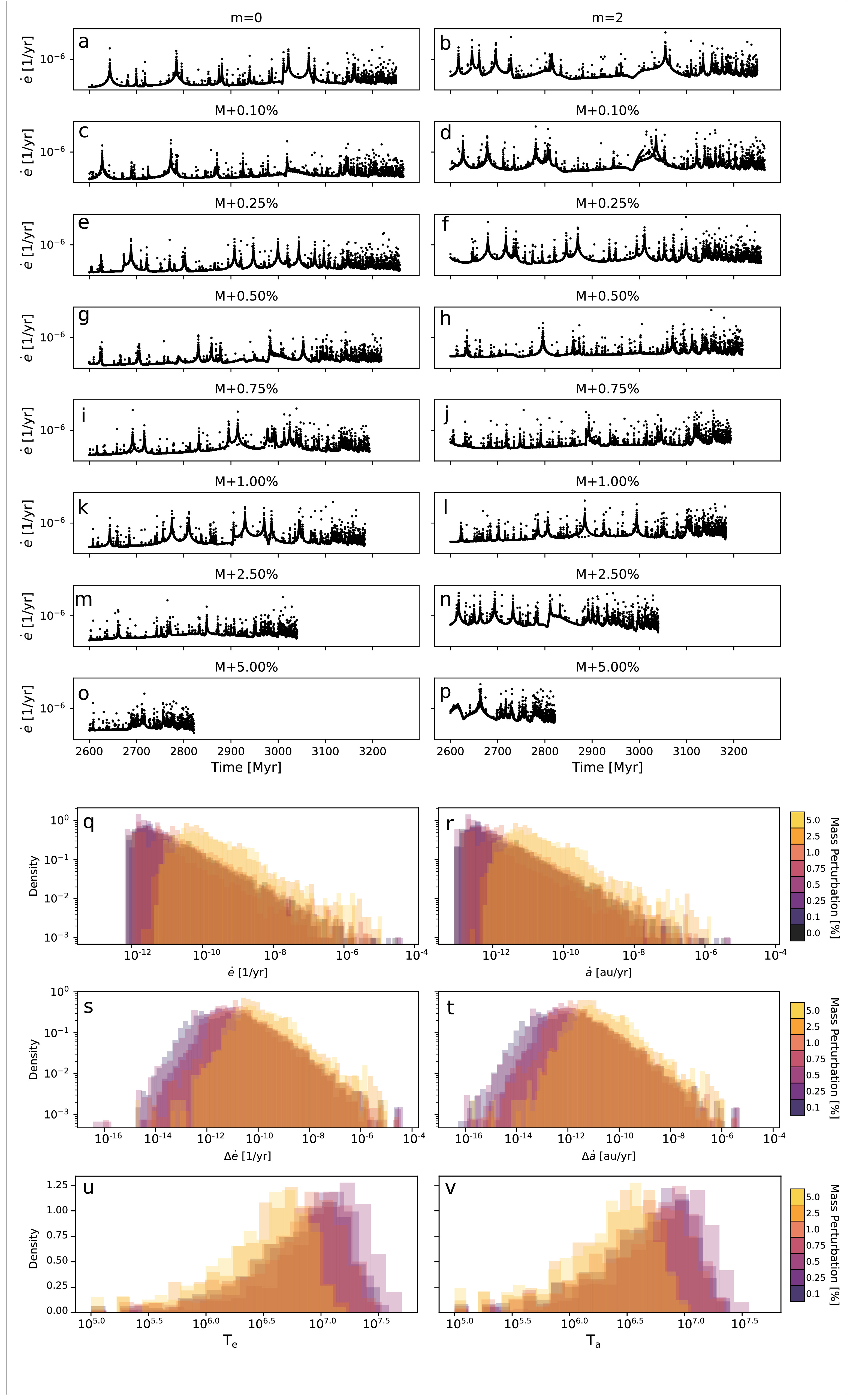}\\
\textbf{Extended Data Figure 5. Limits of tidal evolution predictability for small stellar perturbations.} (a-p) Rate-change of orbital eccentricity from 2.6 Gyr to the end of the main sequence for a $Z=0.02262$ star with a range of mass perturbations, using a base mass of $M=1.36 M_\odot$, decomposed into $m=0$ and $m=2$ modes. The orbital evolution rate is calculated with a fixed-orbit tidal perturber with $e=0.575$, $a=0.153$ and a pseudo synchronous stellar rotation rate. (q) Distribution of $\dot{e}$ and (r) distribution of $\dot{a}$ for each mass perturbation in (a-p). (s) Distribution of $\Delta \dot{e}$ and (t) distribution of $\Delta\dot{a}$. (u) Distribution of time needed to accumulate a $1\sigma$ deviation in $e$ and (v) distribution of time needed to accumulate a $1\sigma$ deviation in $a$. The different colors of distributions are given by the colorbar and denote the perturbations in stellar mass.

\includegraphics[scale=0.1]{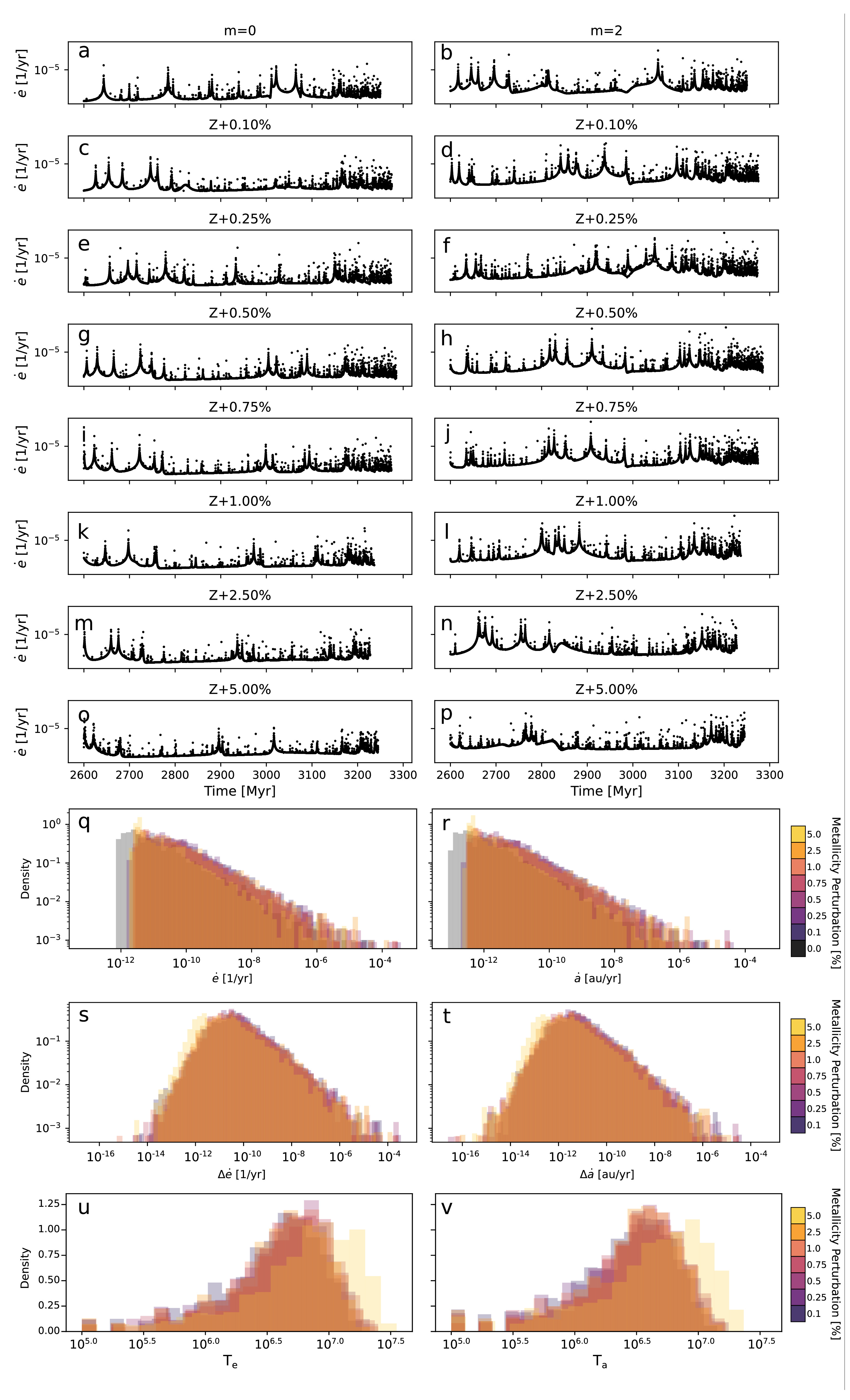}\\
\textbf{Extended Data Figure 6. Limits of tidal evolution predictability for small stellar perturbations.} 
(a-p) Rate-change of orbital eccentricity from 2.6 Gyr to the end of the main sequence for a $M=1.36 M_\odot$ star with a range of metallicity perturbations, using a base metallicity of $Z=0.02262$, decomposed into $m=0$ and $m=2$ modes. The orbital evolution rate is calculated with a fixed-orbit tidal perturber with $e=0.575$, $a=0.153$ and a pseudo synchronous stellar rotation rate. (q) Distribution of $\dot{e}$ and (r) distribution of $\dot{a}$ for each metallicity perturbation in (a-p). (s) Distribution of $\Delta \dot{e}$ and (t) distribution of $\Delta\dot{a}$. (u) Distribution of time needed to accumulate a $1\sigma$ deviation in $e$ and (v) distribution of time needed to accumulate a $1\sigma$ deviation in $a$. The different colors of distributions are given by the colorbar and denote the perturbations in stellar metallicity.

\includegraphics[scale=0.075]{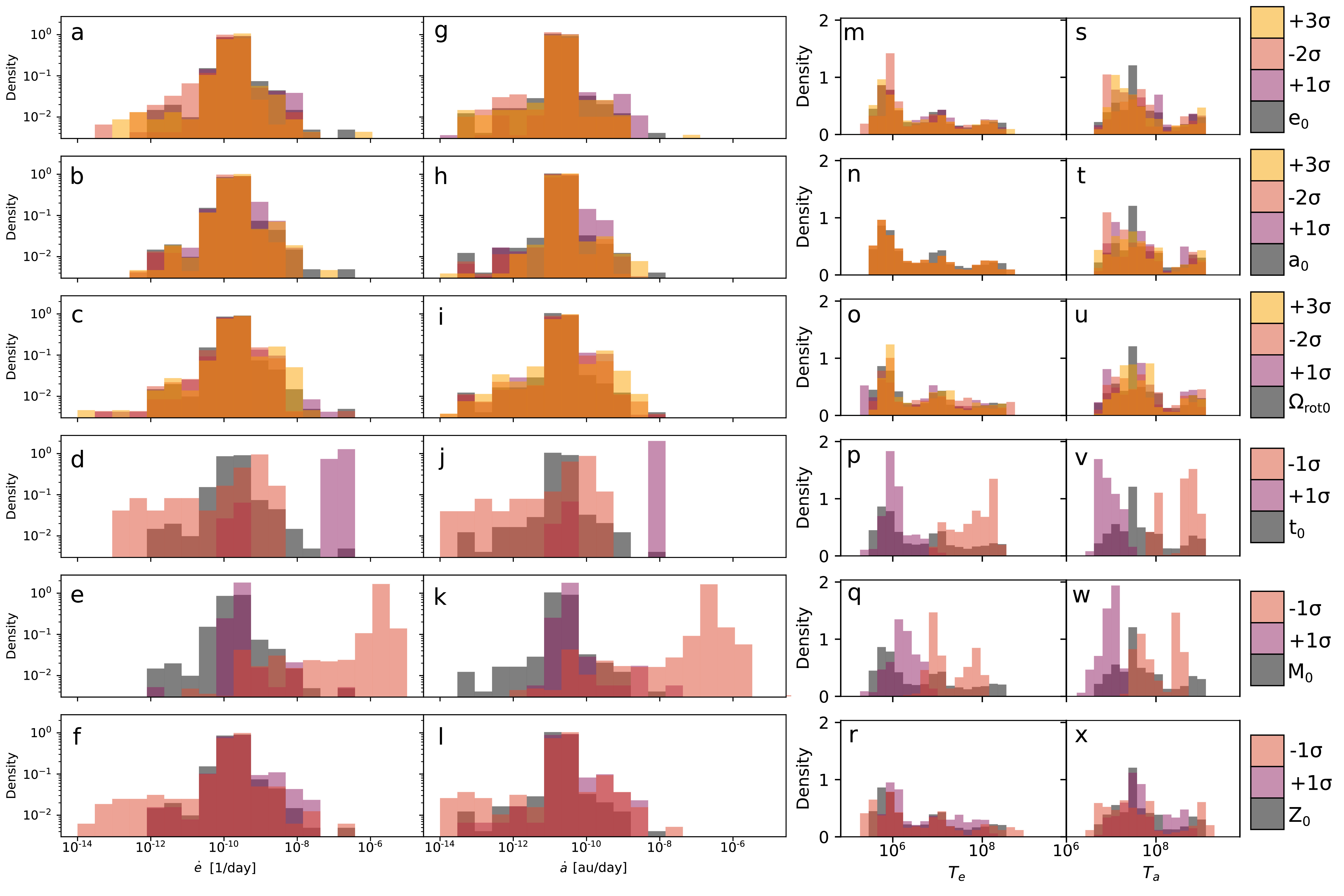}\\
\textbf{Extended Data Figure 7. Limits of tidal evolution predictability within observational bounds.} (a-f) Distribution of $\dot{e}$ and (g-l) distribution of $\dot{a}$ corresponding to the orbital evolution trajectories of fig. 4b in the main text. The panels follow fig. 4b with trajectories differing by perturbations to the initial eccentricity $e_0$, semi-major axis $a_0$, stellar rotation rate $\Omega_{rot0}$, stellar age $t_0$, stellar mass $M_0$, and stellar metallicity $Z_0$. The distribution of time needed to accumulate a $1 \sigma$ deviation in $e$ is given by $T_e$ (m-r) and the distribution of time needed to accumulate a $1 \sigma$ deviation in $a$ is given by $T_a$ (s-x). The panels follow fig. 4b with trajectories differing by perturbations to the initial stellar and orbital parameters. The colors are given by the legend on the right and denote perturbations to the stellar and orbital parameters. 

\newpage

\section*{References}
\bibliography{References.bib}

\begin{thebibliography}{10}
\expandafter\ifx\csname url\endcsname\relax
  \def\url#1{\texttt{#1}}\fi
\expandafter\ifx\csname urlprefix\endcsname\relax\def\urlprefix{URL }\fi
\providecommand{\bibinfo}[2]{#2}
\providecommand{\eprint}[2][]{\url{#2}}

\bibitem{mayor1995jupiter}
\bibinfo{author}{Mayor, M.} \& \bibinfo{author}{Queloz, D.}
\newblock \bibinfo{title}{A jupiter-mass companion to a solar-type star}.
\newblock \emph{\bibinfo{journal}{Nature}} \textbf{\bibinfo{volume}{378}},
  \bibinfo{pages}{355--359} (\bibinfo{year}{1995}).

\bibitem{batygin2016situ}
\bibinfo{author}{Batygin, K.}, \bibinfo{author}{Bodenheimer, P.~H.} \&
  \bibinfo{author}{Laughlin, G.~P.}
\newblock \bibinfo{title}{In situ formation and dynamical evolution of hot
  jupiter systems}.
\newblock \emph{\bibinfo{journal}{\apj}} \textbf{\bibinfo{volume}{829}},
  \bibinfo{pages}{114} (\bibinfo{year}{2016}).

\bibitem{goldreich1980disk}
\bibinfo{author}{Goldreich, P.} \& \bibinfo{author}{Tremaine, S.}
\newblock \bibinfo{title}{Disk-satellite interactions}.
\newblock \emph{\bibinfo{journal}{\apj}} \textbf{\bibinfo{volume}{241}},
  \bibinfo{pages}{425--441} (\bibinfo{year}{1980}).

\bibitem{lin1986tidal}
\bibinfo{author}{Lin, D.~N.} \& \bibinfo{author}{Papaloizou, J.}
\newblock \bibinfo{title}{On the tidal interaction between protoplanets and the
  protoplanetary disk. iii-orbital migration of protoplanets}.
\newblock \emph{\bibinfo{journal}{\apj}} \textbf{\bibinfo{volume}{309}},
  \bibinfo{pages}{846--857} (\bibinfo{year}{1986}).

\bibitem{rasio1996dynamical}
\bibinfo{author}{Rasio, F.~A.} \& \bibinfo{author}{Ford, E.~B.}
\newblock \bibinfo{title}{Dynamical instabilities and the formation of
  extrasolar planetary systems}.
\newblock \emph{\bibinfo{journal}{Science}} \textbf{\bibinfo{volume}{274}},
  \bibinfo{pages}{954--956} (\bibinfo{year}{1996}).

\bibitem{kozai1962secular}
\bibinfo{author}{Kozai, Y.}
\newblock \bibinfo{title}{Secular perturbations of asteroids with high
  inclination and eccentricity}.
\newblock \emph{\bibinfo{journal}{\aj}} \textbf{\bibinfo{volume}{67}},
  \bibinfo{pages}{591--598} (\bibinfo{year}{1962}).

\bibitem{lidov1962evolution}
\bibinfo{author}{Lidov, M.~L.}
\newblock \bibinfo{title}{The evolution of orbits of artificial satellites of
  planets under the action of gravitational perturbations of external bodies}.
\newblock \emph{\bibinfo{journal}{Planetary and Space Science}}
  \textbf{\bibinfo{volume}{9}}, \bibinfo{pages}{719--759}
  (\bibinfo{year}{1962}).

\bibitem{wu2011secular}
\bibinfo{author}{Wu, Y.} \& \bibinfo{author}{Lithwick, Y.}
\newblock \bibinfo{title}{Secular chaos and the production of hot jupiters}.
\newblock \emph{\bibinfo{journal}{\apj}} \textbf{\bibinfo{volume}{735}},
  \bibinfo{pages}{109} (\bibinfo{year}{2011}).

\bibitem{barker2009tidal}
\bibinfo{author}{Barker, A.~J.} \& \bibinfo{author}{Ogilvie, G.~I.}
\newblock \bibinfo{title}{On the tidal evolution of hot jupiters on inclined
  orbits}.
\newblock \emph{\bibinfo{journal}{\mnras}} \textbf{\bibinfo{volume}{395}},
  \bibinfo{pages}{2268--2287} (\bibinfo{year}{2009}).

\bibitem{dawson2018origins}
\bibinfo{author}{Dawson, R.~I.} \& \bibinfo{author}{Johnson, J.~A.}
\newblock \bibinfo{title}{Origins of hot jupiters}.
\newblock \emph{\bibinfo{journal}{\araa}} \textbf{\bibinfo{volume}{56}},
  \bibinfo{pages}{175--221} (\bibinfo{year}{2018}).

\bibitem{eggleton1998equilibrium}
\bibinfo{author}{Eggleton, P.~P.}, \bibinfo{author}{Kiseleva, L.~G.} \&
  \bibinfo{author}{Hut, P.}
\newblock \bibinfo{title}{The equilibrium tide model for tidal friction}.
\newblock \emph{\bibinfo{journal}{\apj}} \textbf{\bibinfo{volume}{499}},
  \bibinfo{pages}{853} (\bibinfo{year}{1998}).

\bibitem{jackson2008tidal}
\bibinfo{author}{Jackson, B.}, \bibinfo{author}{Greenberg, R.} \&
  \bibinfo{author}{Barnes, R.}
\newblock \bibinfo{title}{Tidal evolution of close-in extrasolar planets}.
\newblock \emph{\bibinfo{journal}{\apj}} \textbf{\bibinfo{volume}{678}},
  \bibinfo{pages}{1396} (\bibinfo{year}{2008}).

\bibitem{fuller2017accelerated}
\bibinfo{author}{Fuller, J.}, \bibinfo{author}{Hambleton, K.},
  \bibinfo{author}{Shporer, A.}, \bibinfo{author}{Isaacson, H.} \&
  \bibinfo{author}{Thompson, S.}
\newblock \bibinfo{title}{Accelerated tidal circularization via resonance
  locking in kic 8164262}.
\newblock \emph{\bibinfo{journal}{\mnras: Letters}}
  \textbf{\bibinfo{volume}{472}}, \bibinfo{pages}{L25--L29}
  (\bibinfo{year}{2017}).

\bibitem{hambleton2018kic}
\bibinfo{author}{Hambleton, K.} \emph{et~al.}
\newblock \bibinfo{title}{Kic 8164262: a heartbeat star showing tidally induced
  pulsations with resonant locking}.
\newblock \emph{\bibinfo{journal}{\mnras}} \textbf{\bibinfo{volume}{473}},
  \bibinfo{pages}{5165--5176} (\bibinfo{year}{2018}).

\bibitem{bonomo2017gaps}
\bibinfo{author}{Bonomo, A.~S.} \emph{et~al.}
\newblock \bibinfo{title}{The gaps programme with harps-n at tng-xiv.
  investigating giant planet migration history via improved eccentricity and
  mass determination for 231 transiting planets}.
\newblock \emph{\bibinfo{journal}{\aap}} \textbf{\bibinfo{volume}{602}},
  \bibinfo{pages}{A107} (\bibinfo{year}{2017}).

\bibitem{bakos2007hd}
\bibinfo{author}{Bakos, G.} \emph{et~al.}
\newblock \bibinfo{title}{Hd 147506b: a supermassive planet in an eccentric
  orbit transiting a bright star}.
\newblock \emph{\bibinfo{journal}{\apj}} \textbf{\bibinfo{volume}{670}},
  \bibinfo{pages}{826} (\bibinfo{year}{2007}).

\bibitem{de2017planet}
\bibinfo{author}{De~Wit, J.} \emph{et~al.}
\newblock \bibinfo{title}{Planet-induced stellar pulsations in hat-p-2's
  eccentric system}.
\newblock \emph{\bibinfo{journal}{\apjl}} \textbf{\bibinfo{volume}{836}},
  \bibinfo{pages}{L17} (\bibinfo{year}{2017}).

\bibitem{de2023revisiting}
\bibinfo{author}{de~Beurs, Z.~L.} \emph{et~al.}
\newblock \bibinfo{title}{Revisiting orbital evolution in hat-p-2 b and
  confirmation of hat-p-2 c}.
\newblock \emph{\bibinfo{journal}{\aj}} \textbf{\bibinfo{volume}{166}},
  \bibinfo{pages}{136} (\bibinfo{year}{2023}).

\bibitem{herrero2011wasp}
\bibinfo{author}{Herrero, E.}, \bibinfo{author}{Morales, J.~C.},
  \bibinfo{author}{Ribas, I.} \& \bibinfo{author}{Naves, R.}
\newblock \bibinfo{title}{Wasp-33: the first $\delta$ scuti exoplanet host
  star}.
\newblock \emph{\bibinfo{journal}{\aap}} \textbf{\bibinfo{volume}{526}},
  \bibinfo{pages}{L10} (\bibinfo{year}{2011}).

\bibitem{kalman2022gravity}
\bibinfo{author}{K{\'a}lm{\'a}n, S.} \emph{et~al.}
\newblock \bibinfo{title}{Gravity darkening and tidally perturbed stellar
  pulsation in the misaligned exoplanet system wasp-33}.
\newblock \emph{\bibinfo{journal}{\aap}} \textbf{\bibinfo{volume}{660}},
  \bibinfo{pages}{L2} (\bibinfo{year}{2022}).

\bibitem{kalman2023discovery}
\bibinfo{author}{K{\'a}lm{\'a}n, S.} \emph{et~al.}
\newblock \bibinfo{title}{Discovery of a substellar companion in the tess light
  curve of the $\delta$ scuti$\backslash \gamma$ doradus hybrid pulsator hd
  31221}.
\newblock \emph{\bibinfo{journal}{arXiv preprint arXiv:2305.04000}}
  (\bibinfo{year}{2023}).

\bibitem{de2023engulf}
\bibinfo{author}{De, K.} \emph{et~al.}
\newblock \bibinfo{title}{An infrared transient from a star engulfing a
  planet}.
\newblock \emph{\bibinfo{journal}{Nature}} \textbf{\bibinfo{volume}{617}},
  \bibinfo{pages}{55--60} (\bibinfo{year}{2023}).

\bibitem{zahn1977tidal}
\bibinfo{author}{Zahn, J.-P.}
\newblock \bibinfo{title}{Tidal friction in close binary stars}.
\newblock \emph{\bibinfo{journal}{\aap}} \textbf{\bibinfo{volume}{57}},
  \bibinfo{pages}{383--394} (\bibinfo{year}{1977}).

\bibitem{hut1981tidal}
\bibinfo{author}{Hut, P.}
\newblock \bibinfo{title}{Tidal evolution in close binary systems}.
\newblock \emph{\bibinfo{journal}{Astronomy and Astrophysics, vol. 99, no. 1,
  June 1981, p. 126-140.}} \textbf{\bibinfo{volume}{99}},
  \bibinfo{pages}{126--140} (\bibinfo{year}{1981}).

\bibitem{cowling1941non}
\bibinfo{author}{Cowling, T.~G.}
\newblock \bibinfo{title}{The non-radial oscillations of polytropic stars}.
\newblock \emph{\bibinfo{journal}{\mnras}} \textbf{\bibinfo{volume}{101}},
  \bibinfo{pages}{367} (\bibinfo{year}{1941}).

\bibitem{goodman1998dynamical}
\bibinfo{author}{Goodman, J.} \& \bibinfo{author}{Dickson, E.~S.}
\newblock \bibinfo{title}{Dynamical tide in solar-type binaries}.
\newblock \emph{\bibinfo{journal}{\apj}} \textbf{\bibinfo{volume}{507}},
  \bibinfo{pages}{938} (\bibinfo{year}{1998}).

\bibitem{ogilvie2007tidal}
\bibinfo{author}{Ogilvie, G.} \& \bibinfo{author}{Lin, D.}
\newblock \bibinfo{title}{Tidal dissipation in rotating solar-type stars}.
\newblock \emph{\bibinfo{journal}{\apj}} \textbf{\bibinfo{volume}{661}},
  \bibinfo{pages}{1180} (\bibinfo{year}{2007}).

\bibitem{barker2022tidal}
\bibinfo{author}{Barker, A.~J.}
\newblock \bibinfo{title}{Tidal dissipation due to inertial waves can explain
  the circularization periods of solar-type binaries}.
\newblock \emph{\bibinfo{journal}{\apjl}} \textbf{\bibinfo{volume}{927}},
  \bibinfo{pages}{L36} (\bibinfo{year}{2022}).

\bibitem{weinberg2012nonlinear}
\bibinfo{author}{Weinberg, N.~N.}, \bibinfo{author}{Arras, P.},
  \bibinfo{author}{Quataert, E.} \& \bibinfo{author}{Burkart, J.}
\newblock \bibinfo{title}{Nonlinear tides in close binary systems}.
\newblock \emph{\bibinfo{journal}{\apj}} \textbf{\bibinfo{volume}{751}},
  \bibinfo{pages}{136} (\bibinfo{year}{2012}).

\bibitem{weinberg2017tidal}
\bibinfo{author}{Weinberg, N.~N.}, \bibinfo{author}{Sun, M.},
  \bibinfo{author}{Arras, P.} \& \bibinfo{author}{Essick, R.}
\newblock \bibinfo{title}{Tidal dissipation in wasp-12}.
\newblock \emph{\bibinfo{journal}{\apjl}} \textbf{\bibinfo{volume}{849}},
  \bibinfo{pages}{L11} (\bibinfo{year}{2017}).

\bibitem{macleod2022tidal}
\bibinfo{author}{MacLeod, M.}, \bibinfo{author}{Vick, M.} \&
  \bibinfo{author}{Loeb, A.}
\newblock \bibinfo{title}{Tidal wave breaking in the eccentric lead-in to mass
  transfer and common envelope phases}.
\newblock \emph{\bibinfo{journal}{\apj}} \textbf{\bibinfo{volume}{937}},
  \bibinfo{pages}{37} (\bibinfo{year}{2022}).

\bibitem{goldreich1966q}
\bibinfo{author}{Goldreich, P.} \& \bibinfo{author}{Soter, S.}
\newblock \bibinfo{title}{Q in the solar system}.
\newblock \emph{\bibinfo{journal}{Icarus}} \textbf{\bibinfo{volume}{5}},
  \bibinfo{pages}{375--389} (\bibinfo{year}{1966}).

\bibitem{ogilvie2004tidal}
\bibinfo{author}{Ogilvie, G.~I.} \& \bibinfo{author}{Lin, D.}
\newblock \bibinfo{title}{Tidal dissipation in rotating giant planets}.
\newblock \emph{\bibinfo{journal}{\apj}} \textbf{\bibinfo{volume}{610}},
  \bibinfo{pages}{477} (\bibinfo{year}{2004}).

\bibitem{witte1999tidal}
\bibinfo{author}{Witte, M.} \& \bibinfo{author}{Savonije, G.}
\newblock \bibinfo{title}{Tidal evolution of eccentric orbits in massive binary
  systems; a study of resonance locking}.
\newblock \emph{\bibinfo{journal}{arXiv preprint astro-ph/9909073}}
  (\bibinfo{year}{1999}).

\bibitem{sun2023tides}
\bibinfo{author}{Sun, M.}, \bibinfo{author}{Townsend, R.} \&
  \bibinfo{author}{Guo, Z.}
\newblock \bibinfo{title}{gyre\_tides: Modeling binary tides within the gyre
  stellar oscillation code}.
\newblock \emph{\bibinfo{journal}{\apj}} \textbf{\bibinfo{volume}{945}},
  \bibinfo{pages}{43} (\bibinfo{year}{2023}).

\bibitem{willems2003nonadiabatic}
\bibinfo{author}{Willems, B.}, \bibinfo{author}{Van~Hoolst, T.} \&
  \bibinfo{author}{Smeyers, P.}
\newblock \bibinfo{title}{Nonadiabatic resonant dynamic tides and orbital
  evolution in close binaries}.
\newblock \emph{\bibinfo{journal}{\aap}} \textbf{\bibinfo{volume}{397}},
  \bibinfo{pages}{973--985} (\bibinfo{year}{2003}).

\bibitem{fuller2021inverse}
\bibinfo{author}{Fuller, J.}
\newblock \bibinfo{title}{Inverse tides in pulsating binary stars}.
\newblock \emph{\bibinfo{journal}{\mnras}} \textbf{\bibinfo{volume}{501}},
  \bibinfo{pages}{483--490} (\bibinfo{year}{2021}).

\bibitem{witte2002orbital}
\bibinfo{author}{Witte, M.} \& \bibinfo{author}{Savonije, G.}
\newblock \bibinfo{title}{Orbital evolution by dynamical tides in solar type
  stars-application to binary stars and planetary orbits}.
\newblock \emph{\bibinfo{journal}{\aap}} \textbf{\bibinfo{volume}{386}},
  \bibinfo{pages}{222--236} (\bibinfo{year}{2002}).

\bibitem{ivanov2004tidal}
\bibinfo{author}{Ivanov, P.} \& \bibinfo{author}{Papaloizou, J.}
\newblock \bibinfo{title}{On the tidal interaction of massive extrasolar
  planets on highly eccentric orbits}.
\newblock \emph{\bibinfo{journal}{\mnras}} \textbf{\bibinfo{volume}{347}},
  \bibinfo{pages}{437--453} (\bibinfo{year}{2004}).

\bibitem{wu2018diffusive}
\bibinfo{author}{Wu, Y.}
\newblock \bibinfo{title}{Diffusive tidal evolution for migrating hot
  jupiters}.
\newblock \emph{\bibinfo{journal}{\aj}} \textbf{\bibinfo{volume}{155}},
  \bibinfo{pages}{118} (\bibinfo{year}{2018}).

\bibitem{veras2019tidal}
\bibinfo{author}{Veras, D.} \& \bibinfo{author}{Fuller, J.}
\newblock \bibinfo{title}{Tidal circularization of gaseous planets orbiting
  white dwarfs}.
\newblock \emph{\bibinfo{journal}{\mnras}} \textbf{\bibinfo{volume}{489}},
  \bibinfo{pages}{2941--2953} (\bibinfo{year}{2019}).

\bibitem{vick2019chaotic}
\bibinfo{author}{Vick, M.}, \bibinfo{author}{Lai, D.} \&
  \bibinfo{author}{Anderson, K.~R.}
\newblock \bibinfo{title}{Chaotic tides in migrating gas giants: forming hot
  and transient warm jupiters via lidov--kozai migration}.
\newblock \emph{\bibinfo{journal}{\mnras}} \textbf{\bibinfo{volume}{484}},
  \bibinfo{pages}{5645--5668} (\bibinfo{year}{2019}).

\bibitem{millholland2019obliquity}
\bibinfo{author}{Millholland, S.} \& \bibinfo{author}{Laughlin, G.}
\newblock \bibinfo{title}{Obliquity-driven sculpting of exoplanetary systems}.
\newblock \emph{\bibinfo{journal}{\nata}} \textbf{\bibinfo{volume}{3}},
  \bibinfo{pages}{424--433} (\bibinfo{year}{2019}).

\bibitem{alexander2023orbital}
\bibinfo{author}{Alexander, M.~E.}
\newblock \bibinfo{title}{Orbital precession in short-period hot jupiter
  exoplanet systems}.
\newblock \emph{\bibinfo{journal}{\mnras}} \textbf{\bibinfo{volume}{522}},
  \bibinfo{pages}{1968--1986} (\bibinfo{year}{2023}).

\bibitem{burkart2014dynamical}
\bibinfo{author}{Burkart, J.}, \bibinfo{author}{Quataert, E.} \&
  \bibinfo{author}{Arras, P.}
\newblock \bibinfo{title}{Dynamical resonance locking in tidally interacting
  binary systems}.
\newblock \emph{\bibinfo{journal}{\mnras}} \textbf{\bibinfo{volume}{443}},
  \bibinfo{pages}{2957--2973} (\bibinfo{year}{2014}).

\bibitem{ma2021orbital}
\bibinfo{author}{Ma, L.} \& \bibinfo{author}{Fuller, J.}
\newblock \bibinfo{title}{Orbital decay of short-period exoplanets via tidal
  resonance locking}.
\newblock \emph{\bibinfo{journal}{\apj}} \textbf{\bibinfo{volume}{918}},
  \bibinfo{pages}{16} (\bibinfo{year}{2021}).

\bibitem{fuller2012dynamical}
\bibinfo{author}{Fuller, J.} \& \bibinfo{author}{Lai, D.}
\newblock \bibinfo{title}{Dynamical tides in eccentric binaries and tidally
  excited stellar pulsations in kepler koi-54}.
\newblock \emph{\bibinfo{journal}{\mnras}} \textbf{\bibinfo{volume}{420}},
  \bibinfo{pages}{3126--3138} (\bibinfo{year}{2012}).

\bibitem{burkart2012tidal}
\bibinfo{author}{Burkart, J.}, \bibinfo{author}{Quataert, E.},
  \bibinfo{author}{Arras, P.} \& \bibinfo{author}{Weinberg, N.~N.}
\newblock \bibinfo{title}{Tidal asteroseismology: Kepler’s koi-54}.
\newblock \emph{\bibinfo{journal}{\mnras}} \textbf{\bibinfo{volume}{421}},
  \bibinfo{pages}{983--1006} (\bibinfo{year}{2012}).

\bibitem{burkart2013tidal}
\bibinfo{author}{Burkart, J.}, \bibinfo{author}{Quataert, E.},
  \bibinfo{author}{Arras, P.} \& \bibinfo{author}{Weinberg, N.~N.}
\newblock \bibinfo{title}{Tidal resonance locks in inspiraling white dwarf
  binaries}.
\newblock \emph{\bibinfo{journal}{\mnras}} \textbf{\bibinfo{volume}{433}},
  \bibinfo{pages}{332--352} (\bibinfo{year}{2013}).

\bibitem{barker2010internal}
\bibinfo{author}{Barker, A.~J.} \& \bibinfo{author}{Ogilvie, G.~I.}
\newblock \bibinfo{title}{On internal wave breaking and tidal dissipation near
  the centre of a solar-type star}.
\newblock \emph{\bibinfo{journal}{\mnras}} \textbf{\bibinfo{volume}{404}},
  \bibinfo{pages}{1849--1868} (\bibinfo{year}{2010}).

\bibitem{fuller2016resonance}
\bibinfo{author}{Fuller, J.}, \bibinfo{author}{Luan, J.} \&
  \bibinfo{author}{Quataert, E.}
\newblock \bibinfo{title}{Resonance locking as the source of rapid tidal
  migration in the jupiter and saturn moon systems}.
\newblock \emph{\bibinfo{journal}{\mnras}} \textbf{\bibinfo{volume}{458}},
  \bibinfo{pages}{3867--3879} (\bibinfo{year}{2016}).

\bibitem{lainey2020resonance}
\bibinfo{author}{Lainey, V.} \emph{et~al.}
\newblock \bibinfo{title}{Resonance locking in giant planets indicated by the
  rapid orbital expansion of titan}.
\newblock \emph{\bibinfo{journal}{\nata}} \textbf{\bibinfo{volume}{4}},
  \bibinfo{pages}{1053--1058} (\bibinfo{year}{2020}).

\bibitem{paxton2010modules}
\bibinfo{author}{Paxton, B.} \emph{et~al.}
\newblock \bibinfo{title}{Modules for experiments in stellar astrophysics
  (mesa)}.
\newblock \emph{\bibinfo{journal}{\apjs}} \textbf{\bibinfo{volume}{192}},
  \bibinfo{pages}{3} (\bibinfo{year}{2010}).

\bibitem{paxton2013modules}
\bibinfo{author}{Paxton, B.} \emph{et~al.}
\newblock \bibinfo{title}{Modules for experiments in stellar astrophysics
  (mesa): planets, oscillations, rotation, and massive stars}.
\newblock \emph{\bibinfo{journal}{\apjs}} \textbf{\bibinfo{volume}{208}},
  \bibinfo{pages}{4} (\bibinfo{year}{2013}).

\bibitem{paxton2015modules}
\bibinfo{author}{Paxton, B.} \emph{et~al.}
\newblock \bibinfo{title}{Modules for experiments in stellar astrophysics
  (mesa): binaries, pulsations, and explosions}.
\newblock \emph{\bibinfo{journal}{\apjs}} \textbf{\bibinfo{volume}{220}},
  \bibinfo{pages}{15} (\bibinfo{year}{2015}).

\bibitem{paxton2018modules}
\bibinfo{author}{Paxton, B.} \emph{et~al.}
\newblock \bibinfo{title}{Modules for experiments in stellar astrophysics ():
  Convective boundaries, element diffusion, and massive star explosions}.
\newblock \emph{\bibinfo{journal}{\apjs}} \textbf{\bibinfo{volume}{234}},
  \bibinfo{pages}{34} (\bibinfo{year}{2018}).

\bibitem{paxton2019modules}
\bibinfo{author}{Paxton, B.} \emph{et~al.}
\newblock \bibinfo{title}{Modules for experiments in stellar astrophysics
  (mesa): pulsating variable stars, rotation, convective boundaries, and energy
  conservation}.
\newblock \emph{\bibinfo{journal}{\apjs}} \textbf{\bibinfo{volume}{243}},
  \bibinfo{pages}{10} (\bibinfo{year}{2019}).

\bibitem{jermyn2023modules}
\bibinfo{author}{Jermyn, A.~S.} \emph{et~al.}
\newblock \bibinfo{title}{Modules for experiments in stellar astrophysics
  (mesa): Time-dependent convection, energy conservation, automatic
  differentiation, and infrastructure}.
\newblock \emph{\bibinfo{journal}{\apjs}} \textbf{\bibinfo{volume}{265}},
  \bibinfo{pages}{15} (\bibinfo{year}{2023}).

\bibitem{gossage2023magnetic}
\bibinfo{author}{Gossage, S.}, \bibinfo{author}{Kalogera, V.} \&
  \bibinfo{author}{Sun, M.}
\newblock \bibinfo{title}{Magnetic braking with mesa evolutionary models in the
  single star and low-mass x-ray binary regimes}.
\newblock \emph{\bibinfo{journal}{\apj}} \textbf{\bibinfo{volume}{950}},
  \bibinfo{pages}{27} (\bibinfo{year}{2023}).

\bibitem{townsend2013gyre}
\bibinfo{author}{Townsend, R.} \& \bibinfo{author}{Teitler, S.}
\newblock \bibinfo{title}{Gyre: an open-source stellar oscillation code based
  on a new magnus multiple shooting scheme}.
\newblock \emph{\bibinfo{journal}{\mnras}} \textbf{\bibinfo{volume}{435}},
  \bibinfo{pages}{3406--3418} (\bibinfo{year}{2013}).

\bibitem{townsend2018angular}
\bibinfo{author}{Townsend, R.}, \bibinfo{author}{Goldstein, J.} \&
  \bibinfo{author}{Zweibel, E.}
\newblock \bibinfo{title}{Angular momentum transport by heat-driven g-modes in
  slowly pulsating b stars}.
\newblock \emph{\bibinfo{journal}{\mnras}} \textbf{\bibinfo{volume}{475}},
  \bibinfo{pages}{879--893} (\bibinfo{year}{2018}).

\bibitem{goldstein2020contour}
\bibinfo{author}{Goldstein, J.} \& \bibinfo{author}{Townsend, R.}
\newblock \bibinfo{title}{The contour method: a new approach to finding modes
  of nonadiabatic stellar pulsations}.
\newblock \emph{\bibinfo{journal}{\apj}} \textbf{\bibinfo{volume}{899}},
  \bibinfo{pages}{116} (\bibinfo{year}{2020}).

\bibitem{stassun2019revised}
\bibinfo{author}{Stassun, K.~G.} \emph{et~al.}
\newblock \bibinfo{title}{The revised tess input catalog and candidate target
  list}.
\newblock \emph{\bibinfo{journal}{The Astronomical Journal}}
  \textbf{\bibinfo{volume}{158}}, \bibinfo{pages}{138} (\bibinfo{year}{2019}).

\bibitem{townsend2023discrepant}
\bibinfo{author}{Townsend, R.} \& \bibinfo{author}{Sun, M.}
\newblock \bibinfo{title}{Discrepant approaches to modeling stellar tides and
  the blurring of pseudosynchronization}.
\newblock \emph{\bibinfo{journal}{The Astrophysical Journal}}
  \textbf{\bibinfo{volume}{953}}, \bibinfo{pages}{48} (\bibinfo{year}{2023}).

\bibitem{guo2022new}
\bibinfo{author}{Guo, Z.}, \bibinfo{author}{Ogilvie, G.~I.},
  \bibinfo{author}{Li, G.}, \bibinfo{author}{Townsend, R.~H.} \&
  \bibinfo{author}{Sun, M.}
\newblock \bibinfo{title}{A new window to tidal asteroseismology: non-linearly
  excited stellar eigenmodes and the period spacing pattern in koi-54}.
\newblock \emph{\bibinfo{journal}{\mnras}} \textbf{\bibinfo{volume}{517}},
  \bibinfo{pages}{437--446} (\bibinfo{year}{2022}).

\bibitem{townsend2003semi}
\bibinfo{author}{Townsend, R.}
\newblock \bibinfo{title}{A semi-analytical formula for the light variations
  due to low-frequency g modes in rotating stars}.
\newblock \emph{\bibinfo{journal}{\mnras}} \textbf{\bibinfo{volume}{343}},
  \bibinfo{pages}{125--136} (\bibinfo{year}{2003}).

\bibitem{penoyre2019higher}
\bibinfo{author}{Penoyre, Z.} \& \bibinfo{author}{Sandford, E.}
\newblock \bibinfo{title}{Higher order harmonics in the light curves of
  eccentric planetary systems}.
\newblock \emph{\bibinfo{journal}{\mnras}} \textbf{\bibinfo{volume}{488}},
  \bibinfo{pages}{4181--4194} (\bibinfo{year}{2019}).

\bibitem{prieto2018collection}
\bibinfo{author}{Prieto, C.~A.} \emph{et~al.}
\newblock \bibinfo{title}{A collection of model stellar spectra for spectral
  types b to early-m}.
\newblock \emph{\bibinfo{journal}{\aap}} \textbf{\bibinfo{volume}{618}},
  \bibinfo{pages}{A25} (\bibinfo{year}{2018}).

\bibitem{townsend2023msg}
\bibinfo{author}{Townsend, R.} \& \bibinfo{author}{Lopez, A.}
\newblock \bibinfo{title}{Msg: A software package for interpolating stellar
  spectra in pre-calculated grids}.
\newblock \emph{\bibinfo{journal}{arXiv preprint arXiv:2301.12533}}
  (\bibinfo{year}{2023}).

\bibitem{press1992adaptive}
\bibinfo{author}{Press, W.~H.} \& \bibinfo{author}{Teukolsky, S.~A.}
\newblock \bibinfo{title}{Adaptive stepsize runge-kutta integration}.
\newblock \emph{\bibinfo{journal}{Computers in Physics}}
  \textbf{\bibinfo{volume}{6}}, \bibinfo{pages}{188--191}
  (\bibinfo{year}{1992}).

\bibitem{chewi2021fast}
\bibinfo{author}{Chewi, S.} \emph{et~al.}
\newblock \bibinfo{title}{Fast and smooth interpolation on wasserstein space}.
\newblock In \emph{\bibinfo{booktitle}{International Conference on Artificial
  Intelligence and Statistics}}, \bibinfo{pages}{3061--3069}
  (\bibinfo{organization}{PMLR}, \bibinfo{year}{2021}).

\bibitem{bryan2023capturing}
\bibinfo{author}{Bryan, J.}, \bibinfo{author}{Frank, W.~B.} \&
  \bibinfo{author}{Audet, P.}
\newblock \bibinfo{title}{Capturing seismic velocity changes in receiver
  functions with optimal transport}.
\newblock \emph{\bibinfo{journal}{\gji}} \textbf{\bibinfo{volume}{234}},
  \bibinfo{pages}{1282--1306} (\bibinfo{year}{2023}).

\bibitem{bryan2024zenodo}
\bibinfo{author}{Bryan, J.}, \bibinfo{author}{de~Wit, J.},
  \bibinfo{author}{Sun, M.}, \bibinfo{author}{de~Beurs, Z.} \&
  \bibinfo{author}{Townsend, R.}
\newblock \bibinfo{title}{The coevolution of migrating planets and their
  pulsating stars through episodic resonance locking.} (\bibinfo{year}{2024}).
\newblock \urlprefix\url{http://dx.doi.org/10.5281/zenodo.11509036}.

\end{thebibliography}
\bibliographystyle{naturemag}


\end{methods}
\end{document}


\maketitle
\begin{affiliations}
 \item Department of Earth, Atmospheric and Planetary Sciences, Massachusetts Institute of Technology, Cambridge, MA, USA;
 \item Department of Astronomy, University of Wisconsin -- Madison, Madison, WI, USA;
 \item Center for Interdisciplinary Exploration and Research in Astrophysics, Department of Physics \& Astronomy, Northwestern University, Evanston, IL, USA
\end{affiliations}

\begin{abstract}
\end{abstract}

\flushbottom
\maketitle

\section{Figures and Tables}
\includegraphics[scale=0.8]{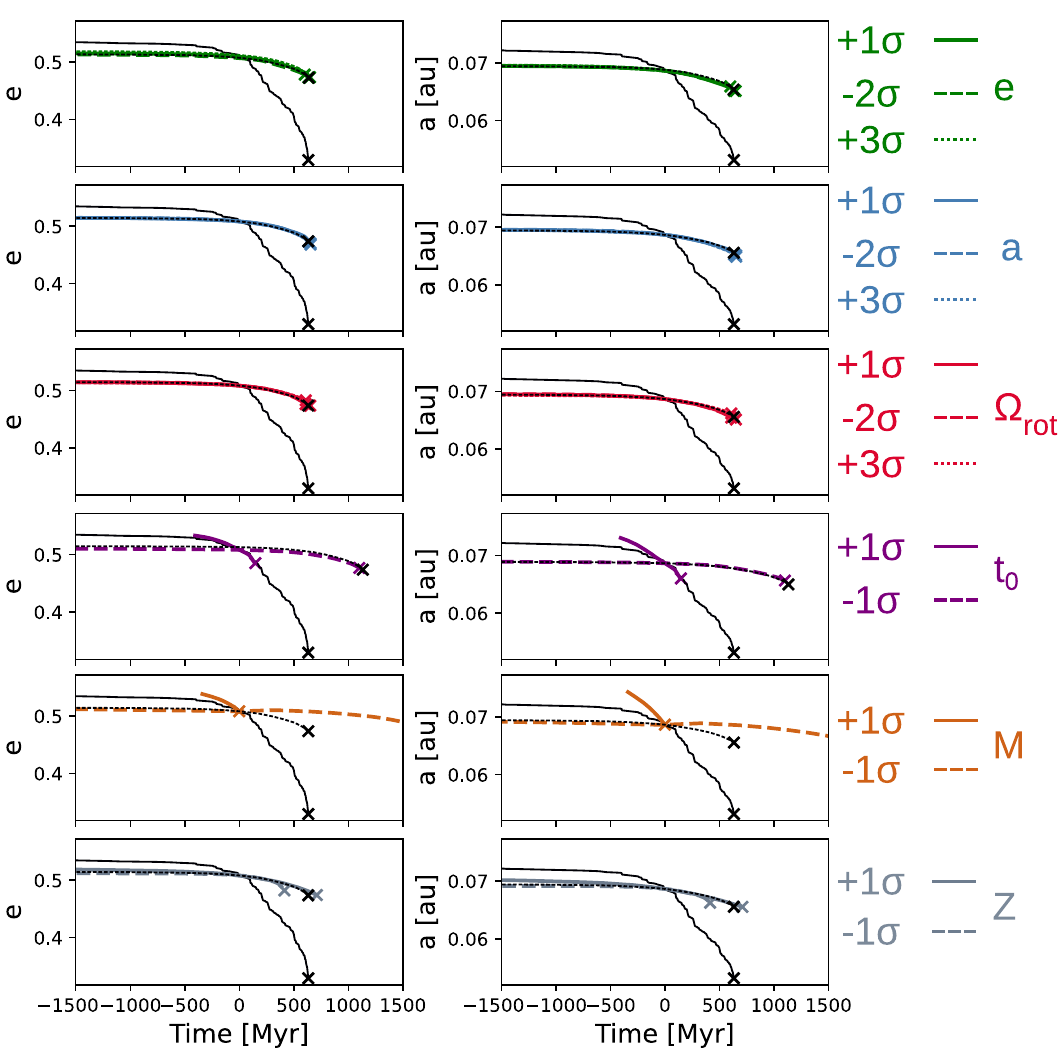}\\
\textbf{Supplementary Figure 1. Comparison of equilibrium and dynamical tidal migration.} Comparison of GYRE-tides models considering only equilibrium tides to GYRE-tides models considering the full tidal response. The solid black line corresponds to the baseline GYRE-tides model with the full tidal response. The dotted black line corresponds to the GYRE-tides model with only the equilibrium tidal response. Colored lines correspond to those in Fig. 4b, but for only the equilibrium tidal response. The baseline $\Omega_\mathrm{rot}$ is set to the value estimated for HAT-P-2 by Ref.\cite{bonomo2017gaps}.

\includegraphics[scale=0.8]{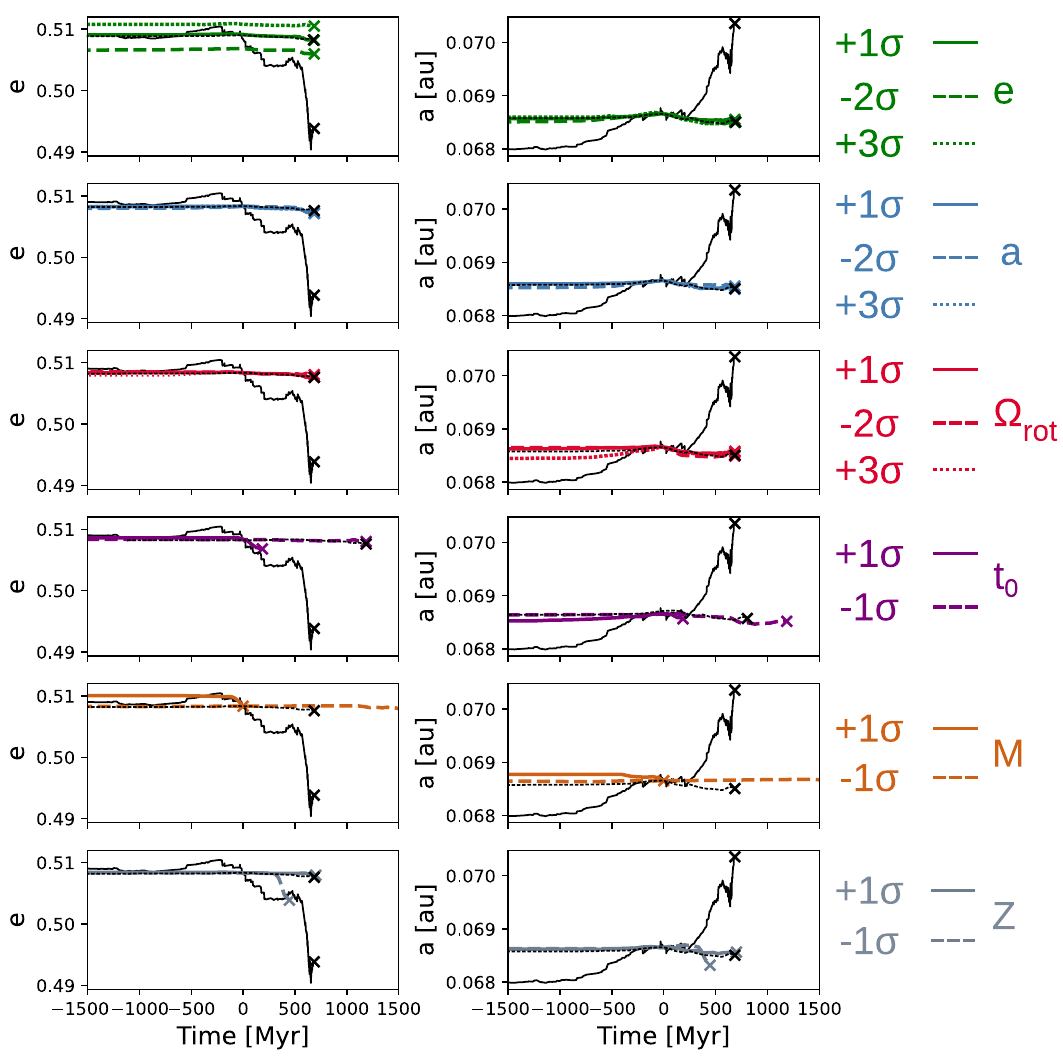}\\
\textbf{Supplementary Figure 2. Comparison of equilibrium and dynamical tidal migration.} Comparison of GYRE-tides models considering only equilibrium tides to GYRE-tides models considering the full tidal response. The solid black line corresponds to the baseline GYRE-tides model with the full tidal response. The dotted black line corresponds to the GYRE-tides model with only the equilibrium tidal response. Colored lines correspond to those in Fig. 4b, but for only the equilibrium tidal response. The layout follows Supplementary Fig. 1. The baseline $\Omega_\mathrm{rot}$ is set to 1.5$\Omega_\mathrm{ps}$, the crossover point for the tidal torque in Ref.\cite{sun2023tides}.

\includegraphics[scale=0.8]{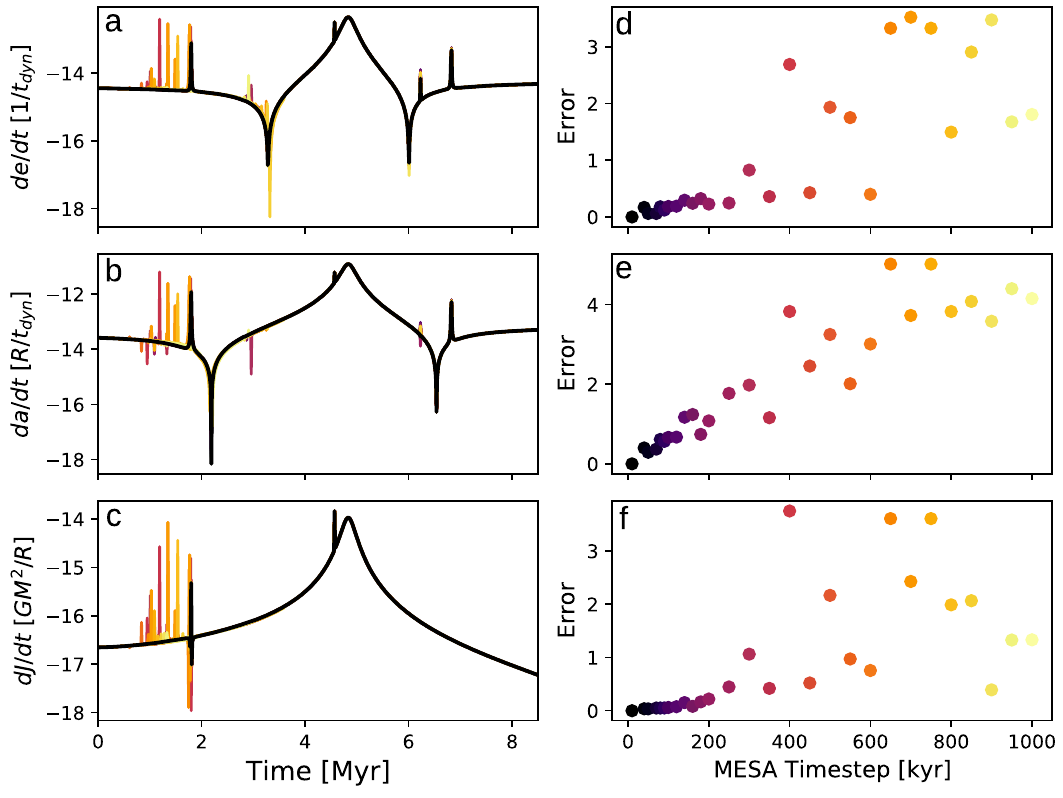}\\
\textbf{Supplementary Figure 3. MESA model interpolation error.} (a-b) Orbital evolution rates, plotted in absolute value and on a log-scale, for eccentricity $e$, semi-major axis $a$, and stellar angular momentum $J$. The orbital configuration is fixed and the star evolves. The ground truth (black) is given by the orbital evolution rates evaluated directly on the stellar profiles produced by MESA ($dt_{gyre}=dt_{mesa}=10^4$ yr). Colored lines correspond to orbital evolution rates evaluated on interpolated stellar models (i.e. $dt_{gyre}<dt_{mesa}$). (d-f) L2 norm between the (log absolute value) orbital evolution rates obtained in the ground truth model and for sparser models. $dt_{mesa}$ is given on the x-axis and $dt_{gyre}$ is fixed at $dt=10^4$ yr. All calculations use the reference properties of the HAT-P-2 system\cite{bonomo2017gaps}.